\documentstyle[preprint,eqsecnum,aps,pra]{revtex} \draft

\def\bra#1{\langle #1 |} \def\ket#1{| #1\rangle}
 \def\inprod#1,#2{\langle
\, #1 \, , \, #2 \, \rangle} \def\braket#1,#2,#3 {\bra{#1} \, #2 \,
\ket{#3} }

\begin{document}

\preprint{DAMTP-96-88, gr-qc/9610028}

\title{Quantum Prediction Algorithms}

\author{Adrian Kent\thanks{E-mail: apak@damtp.cambridge.ac.uk} and Jim
  McElwaine\thanks{E-mail: jnm11@damtp.cambridge.ac.uk}}

\address{Department of Applied Mathematics and Theoretical Physics,\\
  University of Cambridge,\\ Silver Street, Cambridge CB3 9EW, U.K.}

\date{14th October, 1996}
\maketitle\vspace*{-1em}
\begin{abstract}
\noindent
The consistent histories formulation of the quantum theory of a closed
system with pure initial state defines an infinite number of
incompatible consistent sets, each of which gives a possible
description of the physics.  We investigate the possibility of using
the properties of the Schmidt decomposition to define an algorithm
which selects a single, physically natural, consistent set.  We
explain the problems which arise, set out some possible algorithms,
and explain their properties with the aid of simple models.  Though
the discussion is framed in the language of the consistent histories
approach, it is intended to highlight the difficulty in making any
interpretation of quantum theory based on decoherence into a
mathematically precise theory.
\end{abstract}

\vspace{15pt} \pacs{PACS numbers: 03.65.Bz, 98.80.H
  \\[10pt] Submitted to Phys. Rev. A}

\tableofcontents

\section{Introduction}\label{sec:alg:intro} 

It is hard to find an entirely satisfactory interpretation of the
quantum theory of closed systems, since quantum theory does not
distinguish physically interesting time-ordered sequences of
operators.  The consistent histories approach to quantum theory was
originally developed by Griffiths\cite{Griffiths:1},
Omn\`es\cite{Omnes:1}, and Gell-Mann \& Hartle\cite{CEPI:GMH}. One of
its virtues, in our view, is that it allows the problems of the
quantum theory of closed systems to be formulated precisely enough to
allow us to explore possible solutions.  A natural probability
distribution is defined on each consistent set of histories, allowing
probabilistic predictions to be made from the initial data.  There are
infinitely many consistent sets, which are incompatible in the sense
that pairs of sets generally admit no physically sensible joint
probability distribution whose marginal distributions agree with those
on the individual sets.  Indeed the standard
no-local-hidden-variables-theorems show that there is no joint
probability distribution defined on the collection of histories
belonging to all consistent
sets\cite{Goldstein:Page,Dowker:Kent:approach}.  Hence the set
selection problem: probabilistic predictions can only be made
conditional on a choice of consistent set, yet the consistent
histories formalism gives no way of singling out any particular set or
sets as physically interesting.

One possible solution to the set selection problem would be an axiom
which identifies a unique physically interesting set, or perhaps a
class of such sets, from the initial state and the dynamics.  Another
would be the identification of a physically natural measure on the
space of consistent sets, according to which the physically relevant
consistent set is randomly chosen.  No workable solution has yet been
proposed, however.

The problem remains essentially unaltered if the predictions are
conditioned on a large collection of data\cite{Dowker:Kent:approach},
and even if predictions are made conditional on approximately
classical physics being observed\cite{Kent:quasi}. The consistent
histories approach thus violates both standard scientific criteria and
ordinary
intuition\cite{Dowker:Kent:approach,Kent:quasi,Dowker:Kent:properties,%
Kent:bohm,Kent:contra,Kent:implications}.  In our view, the present
version of the consistent histories formalism is too weakly predictive
in almost all plausible physical situations to be considered a
fundamental scientific theory.  Nonetheless, we believe that the
consistent histories approach gives a new way of looking at quantum
theory which raises intriguing questions and should, if possible, be
developed further.

The status of the consistent histories approach remains controversial:
much more optimistic assessments of the present state of the
formalism, can be found, for example, in
refs.\cite{CEPI:GMH,omnesbook,griffithschqr}.  It is, though,
generally agreed that set selection criteria should be investigated.
For if quantum theory correctly describes macroscopic physics then, it
is believed, real world experiments and observations can be described
by what Gell-Mann and Hartle term \emph{quasiclassical} consistent
sets of histories.  Roughly speaking, quasiclassical sets are defined
by projection operators which involve similar variables at different
times and which satisfy classical equations of motion, to a very good
approximation, most of the time.  No precise definition of
quasiclassicality has yet been found, nor is any systematic way known
of identifying quasiclassical sets within any given model or theory.
Whether Gell-Mann and Hartle's program of characterising
quasiclassical sets is taken as a fundamental problem or a
phenomenological one, any solution must clearly involve some sort of
set selection mechanism.

In this paper, we consider one particular line of attack on this
problem: the attempt to select consistent sets by using the Schmidt
decomposition together with criteria intrinsic to the consistent
histories formalism.  The paper is exploratory in spirit: our aims
here are to point out obstacles, raise questions, set out some
possible selection principles, and explain their properties.

Our discussion is framed in the language of the consistent histories
approach to quantum theory, but we believe it is of wider relevance.
Many modern attempts to provide an interpretation of quantum theory
rely, ultimately, on the fact that quantum subsystems decohere.
Subsystems considered include the brains of observers, the pointers of
measuring devices, and abstractly defined subspaces of the total
Hilbert space.  Whichever, the moral is intended to be that
decoherence selects the projection operators, or space-time events, or
algebras of observables which characterise the physics of the
subsystem as it is experienced or observed.  There is no doubt that
understanding the physics of decoherence \emph{does} provide a very
good intuitive grasp of how to identify operators from which our
everyday picture of real-world quasiclassical physics can be
constructed and this lends some support to the hope that a workable
interpretation of quantum theory --- a plausible successor to the
Copenhagen interpretation --- \emph{could} possibly be constructed
along the lines just described.

However, it seems to us that the key question is whether such an
interpretation can be made mathematically precise.\footnote{Even those
who believe that an interpretation relying on intuitive ideas or
verbal prescriptions is acceptable would, we hope, concede that it is
interesting to ask whether those ideas and prescriptions \emph{can} be
set out mathematically.}  That is, given a decohering subsystem, can
we find general rules which precisely specify operators (or other
mathematical objects) which allow us to recover the subsystem's
physics as we experience or observe it?  From this point of view, we
illustrate below how one might go about setting out such rules, and
the sort of problems which arise.

\subsection{Consistent histories} 

We use a version of the consistent histories formalism in which the
initial conditions are defined by a pure state, the basic objects of
the formalism are branch-dependent sets of projections, and
consistency is defined by Gell-Mann and Hartle's decoherence criterion
eq.~(\ref{consist}).

Let $\ket{\psi}$ be the initial state of a quantum system.  A
\emph{branch-dependent set of histories} is a set of products of
projection operators indexed by the variables $\alpha = \{ \alpha_n ,
\alpha_{n-1} , \ldots , \alpha_1 \}$ and corresponding time
coordinates $\{ t_n , \ldots , t_1 \}$, where the ranges of the
$\alpha_k$ and the projections they define depend on the values of
$\alpha_{k-1} , \ldots , \alpha_1 $, and the histories take the form:
\begin{equation} \label{histories}
  C_{\alpha} = P_{\alpha_n}^n (t_n ; \alpha_{n-1} , \ldots , \alpha_1
  ) P_{\alpha_{n-1}}^{n-1} (t_{n-1} ; \alpha_{n-2} , \ldots , \alpha_1
  ) \ldots P_{\alpha_1}^1 ( t_1 )\,.
\end{equation}
Here, for fixed values of $\alpha_{k-1} , \ldots , \alpha_1$, the
$P^k_{\alpha_k} (t_k ; \alpha_{k-1} , \ldots , \alpha_1 )$ define a
projective decomposition of the identity\footnote{For brevity, we
refer to projective decompositions of the identity as projective
decompositions hereafter.}  indexed by $\alpha_k$, so that
$\sum_{\alpha_k} P^k_{\alpha_k} (t_k ; \alpha_{k-1} , \ldots ,
\alpha_1 ) = 1 $ and
\begin{equation} \label{decomp}
P^k_{\alpha_k} (t_k ; \alpha_{k-1} , \ldots , \alpha_1 )
P^k_{\alpha'_k} (t_k ; \alpha_{k-1} , \ldots , \alpha_1 ) =
\delta_{\alpha_k \alpha'_k } P^k_{\alpha_k} (t_k ; \alpha_{k-1} ,
\ldots , \alpha_1 )\,.
\end{equation}
The set of histories is \emph{consistent\footnote{Several different
    consistency/decoherence criteria are used in the literature, all
    of which are believed to be compatible with the standard
    quasiclassical descriptions of realistic physical examples.  This
    particular criterion is generally known as \emph{medium
    consistency} or \emph{medium decoherence}; it will be used
    throughout the paper.}}  if and only if
\begin{equation} \label{consist}
  D_{\alpha\beta} = \langle\psi| C_{\beta}^\dagger C_{\alpha}
  |\psi\rangle = \delta_{\alpha \beta } p ( \alpha ) ,
\end{equation}
in which case $p(\alpha)$ is interpreted as the probability of the
history $\alpha$. $D$ is the decoherence matrix.  Here and later,
though we use the compact notation $\alpha$ to refer to a history, we
intend the individual projection operators and their associated times
to define the history.  The histories of non-zero probability in a
consistent set thus correspond precisely to the non-zero vectors
$C_{\alpha} |\psi\rangle$.  According to the standard view of the
consistent histories formalism, which we adopt here, it is only
consistent sets which are of physical relevance.  The dynamics are
defined purely by the Hamiltonian, with no collapse postulate, but
each projection in the history can be thought of as corresponding to a
historical event, taking place at the relevant time.  If a given
history is realised, its events correspond to extra physical
information, neither deducible from the state vector nor influencing
it.

Most projection operators involve rather obscure physical quantities,
so that it is hard to interpret a general history in familiar
language.  However, given a sensible model, with Hamiltonian and
canonical variables specified, one can construct sets of histories
which describe familiar physics and check that they are indeed
consistent to a very good approximation.  For example, a useful set of
histories for describing the solar system could be defined by
projection operators whose non-zero eigenspaces contain states in
which a given planet's centre of mass is located in a suitably chosen
small volumes of space at the relevant times, and one would expect a
sensible model to show that this is a consistent set and that the
histories of significant probability are those agreeing with the
trajectories predicted by general relativity.

More generally, Gell-Mann and Hartle\cite{CEPI:GMH} introduce the
notion of a \emph{quasiclassical domain}: a consistent set which is
complete --- so that it cannot be non-trivially consistently extended
by more projective decompositions --- and is defined by projection
operators which involve similar variables at different times and which
satisfy classical equations of motion, to a very good approximation,
most of the time.  The notion of a quasiclassical domain seems
natural, though presently imprecisely defined.  Its heuristic
definition is motivated by the familiar example of the hydrodynamic
variables --- densities of chemical species in small volumes of space,
and similar quantities --- which characterise our own quasiclassical
domain.  Here the branch-dependence of the formalism plays an
important role, since the precise choice of variables (most obviously,
the sizes of the small volumes) we use depends on earlier historical
events.  The formation of our galaxy and solar system influences all
subsequent local physics; even present-day quantum experiments have
the potential to do so significantly, if we arrange for large
macroscopic events to depend on their results.

It should be stressed that, according to all the developers of the
consistent histories approach, quasiclassicality and related
properties are interesting notions to study within, not defining
features of, the formalism.  In the view of the formalism's
developers, all consistent sets of histories have the same physical
status, though in any realistic example we are likely to be more
interested in the descriptions of the physics given by some than by
others.

Identifying interesting consistent sets of histories is presently more
of an art than a science.  One of the original aims of the consistent
histories formalism, stressed in particular by Griffiths and Omn\`es,
was to provide a theoretical justification for the intuitive language
often used, both by theorists and experimenters, in analysing
laboratory setups.  Even here, though there are many interesting
examples in the literature of consistent sets which give a natural
description of particular experiments, no general principles have been
found by which such sets can be identified.  Identifying interesting
consistent sets in quantum cosmological models or in real world
cosmology seems to be still harder, although some interesting criteria
stronger than consistency have recently been
proposed\cite{Kent:implications,gmhstrong}.

\subsection{The Schmidt decomposition} 

We consider a closed quantum system with pure initial state vector
$|\psi (0)\rangle$ in a Hilbert space ${\cal{H}}$ with Hamiltonian
$H$.  We suppose that ${\cal{H}} = {\cal{H}}_1 \otimes {\cal{H}}_2$;
we write $\dim ( {\cal{H}}_j ) = d_j$ and we suppose that $d_1 \leq
d_2 < \infty$.  With respect to this splitting of the Hilbert space,
the \emph{Schmidt decomposition} of $|\psi (t) \rangle$ is an
expression of the form
\begin{equation} \label{schmidteqn}
  |\psi (t) \rangle = \sum_{i=1}^{d_1} \, [p_i(t)]^{1/2} \, | w_i
  (t)\rangle_1 \otimes |w_i (t)\rangle_2 \, ,
\end{equation}
where the \emph{Schmidt states} $\{ |w_i\rangle_1 \}$ and $\{
|w_i\rangle_2\}$ form, respectively, an orthonormal basis of
${\cal{H}}_1$ and part of an orthonormal basis of ${\cal{H}}_2$, the
functions $p_i (t)$ are real and positive, and we take the positive
square root.  For fixed time $t$, any decomposition of the form
eq.~(\ref{schmidteqn}) then has the same list of probability weights
$\{ p_i (t) \}$, and the decomposition~(\ref{schmidteqn}) is unique if
these weights are all different. These probability weights are the
eigenvalues of the reduced density matrix.

This simple result, proved by Schmidt in 1907\cite{schmidt}, means
that at any given time there is generically a natural decomposition of
the state vector relative to any fixed split ${\cal{H}} = {\cal{H}}_1
\otimes {\cal{H}}_2$, which defines a basis on the smaller space
${\cal H}_1$ and a partial basis on ${\cal H}_2$.  The decomposition
has an obvious application in standard Copenhagen quantum theory
where, if the two spaces correspond to subsystems undergoing a
measurement-type interaction, it describes the final
outcomes\cite{vonneumann}.

It has more than once been suggested that the Schmidt decomposition
\emph{per se} might define a fundamental interpretation of quantum
theory. According to one line of thought, it defines the structure
required in order to make precise sense of Everett's
ideas\cite{deutsch}.  Another idea which has attracted some attention
is that the Schmidt decomposition itself defines a fundamental
interpretation\cite{dieks,vanf,healey,kochen}.  Some critical comments
on this last program, motivated by its irreconcilability with the
quantum history probabilities defined by the decoherence matrix, can
be found in ref.\cite{Kent:pla}.

Though a detailed critique is beyond our scope here, it seems to us
that any attempt to interpret quantum theory which relies solely on
the properties of the Schmidt decomposition must fail, even if some
fixed choice of ${\cal H}_1$ and ${\cal H}_2$ is allowed.  The Schmidt
decomposition seems inadequate as, although it allows a plausible
interpretation of the quantum state at a single fixed time, its time
evolution has no natural interpretation consistent with the
predictions of Copenhagen quantum theory.

Many studies have been made of the behaviour of the Schmidt
decomposition during system-environment interactions.  In developing
the ideas of this paper, we were influenced in particular by
Albrecht's
investigations\cite{Albrecht:decoherence,Albrecht:collapsing} of the
behaviour of the Schmidt decomposition in random Hamiltonian
interaction models and the description of these models by consistent
histories.

\subsection{Combining consistency and the Schmidt decomposition} 

The idea motivating this paper is that the combination of the ideas of
the consistent histories formalism and the Schmidt decomposition might
allow us to define a mathematically precise and physically interesting
description of the quantum theory of a closed system.  The Schmidt
decomposition defines four natural classes of projection operators,
which we refer to collectively as \emph{Schmidt projections}. These
take the form
\begin{equation} \label{schmidtprojs1}
  \begin{array}{lll}
  P_i^1 (t) = \ket{w_i (t) }_1 \bra{w_i (t) }_1 \otimes I_2
  &\mbox{and}& \overline P^1 = I_1 \otimes I_2 - \sum_i P_i^1 (t)\, ,
  \\ P_i^2 (t) = I_1 \otimes \ket{w_i (t) }_2 \bra{w_i (t) }_2
  &\mbox{and}& \overline P^2 = I_1 \otimes I_2 - \sum_i P_i^2 (t) \, ,
  \\ P_i^{3} (t) = \ket{w_i (t) }_1 \bra{w_i (t) }_1 \otimes \ket{w_i
  (t) }_2 \bra{w_i (t) }_2 &\mbox{and}& \overline P^3 = I_1 \otimes
  I_2 - \sum_i P_i^{3} (t) \, , \\ P_{ij}^{4} (t) = \ket{w_i (t) }_1
  \bra{w_i (t) }_1 \otimes \ket{w_j (t) }_2 \bra{w_j (t) }_2
  &\mbox{and}& \overline P^4 = I_1 \otimes I_2 - \sum_{ij} P_{ij}^{4}
  (t)\,.
 \end{array}
\end{equation}
If $\mbox{dim}{\cal H}_1 = \mbox{dim}{\cal H}_2$ the complementary
projections $ \overline P^1$, $ \overline P^2$ and $\overline P^4$ are
zero.

Since the fundamental problem with the consistent histories approach
seems to be that it allows far too many consistent sets of
projections, and since the Schmidt projections appear to be natural
dynamically determined projections, it seems sensible to explore the
possibility that a physically sensible rule can be found which selects
a consistent set or sets from amongst those defined by Schmidt
projections.

The first problem in implementing this idea is choosing the split
${\cal{H}} = {\cal{H}}_1 \otimes {\cal{H}}_2$.  In analysing
laboratory experiments, one obvious possibility is to separate the
system and apparatus degrees of freedom.  Other possibilities of more
general application are to take the split to correspond to more
fundamental divisions of the degrees of freedom --- fermions and
bosons, or massive and massless particles, or, one might speculate,
the matter and gravitational fields in quantum gravity.  Some such
division would necessarily have to be introduced if this proposal were
applied to cosmological models.

Each of these choices seems interesting to us in context, but none, of
course, is conceptually cost-free.  Assuming a division between system
and apparatus in a laboratory experiment seems to us unacceptable in a
fundamental theory, reintroducing as it does the Heisenberg cut which
post-Copenhagen quantum theory aims to eliminate.  It seems
justifiable, though, for the limited purpose of discussing the
consistent sets which describe physically interesting histories in
laboratory situations.  It also allows useful tests: if an algorithm
fails to give sensible answers here, it should probably be discarded;
if it succeeds, applications elsewhere may be worth exploring.

Postulating a fundamental split of Hilbert space, on the other hand,
seems to us acceptable in principle.  If the split chosen were
reasonably natural, and if it were to produce a well-defined and
physically sensible interpretation of quantum theory applied to closed
systems, we would see no reason not to adopt it.  This seems a
possibility especially worth exploring in quantum cosmology, where any
pointers towards calculations that might give new physical insight
would be welcome.

Here, though, we leave aside these motivations and the conceptual
questions they raise, as there are simpler and more concrete problems
which first need to be addressed.  Our aim in this paper is simply to
explain the problems which arise in trying to define consistent set
selection algorithms using the Schmidt decomposition, to set out some
possibilities, and to explain their properties, using simple models of
quantum systems interacting with an idealised experimental device or
with a series of such devices.

The most basic question here is precisely which of the Schmidt
projections should be used.  Again, our view is pragmatic: we would
happily adopt any choice that gave physically interesting results.
Where we discuss the abstract features of Schmidt projection
algorithms below, the discussion is intended to apply to all four
choices. When we consider simple models of experimental setups, we
take ${\cal{H}}_1$ to describe the system variables and ${\cal{H}}_2$
the apparatus or environment.  Here we look for histories which
describe the evolution of the system state, tracing over the
environment, and so discuss set selection algorithms which use only
the first class of Schmidt projections: the other possibilities are
also interesting, but run into essentially the same problems.  Thus,
in the remainder of the paper, we use the term Schmidt projection to
mean the system space Schmidt projections denoted by $P_i^1$ and
$\overline P^1$ defined in eq.~(\ref{schmidtprojs1}).

In most of the following discussion, we consider algorithms which use
only the properties of the state vector $ |\psi (t) \rangle$ and its
Schmidt decomposition to select a consistent set.  However, we will
also consider later the possibility of reconstructing a branching
structure defined by the decomposition
\begin{equation}
 |\psi (t) \rangle = \sum_{i=1}^{N(t)} |\psi_i (t) \rangle \, ,
\end{equation} 
in which the selected set is branch-dependent and the distinct
orthogonal components $ |\psi_i (t) \rangle$ correspond to the
different branches at time $t$.  In this case, we will consider the
Schmidt decompositions of each of the $ |\psi_i (t) \rangle$
separately.  Again, it will be sufficient to consider only the first
class of Schmidt projections.  In fact, for the branch-dependent
algorithms we consider, all of the classes of Schmidt projection
select the same history vectors and hence select physically equivalent
consistent sets.

\section{Approximate consistency and non-triviality}
\label{sec:approxcon} 

In realistic examples it is generally difficult to find simple
examples of physically interesting sets that are exactly consistent.
For simple physical projections, the off-diagonal terms of the
decoherence matrix typically decay exponentially.  The sets of
histories defined by these projections separated by times much larger
than the decoherence time, are thus typically very nearly but not
precisely
consistent\cite{GM:Hartle:classical,Caldeira:Leggett,Joos:Zeh,%
Dowker:Halliwell,Pohle,Paz:brownian:motion,%
Zurek:preferred:observables,Anastopoulos:Halliwell,%
Tegmark:Shapiro,Paz:Zurek:classicality,Hartle:spacetime,%
Zurek:transition}.  Histories formed from Schmidt projections are no
exception: they give rise to exactly consistent sets only in special
cases, and even in these cases the exact consistency is unstable under
perturbations of the initial conditions or the Hamiltonian.

The lack of simple exactly consistent sets is not generally thought to
be a fundamental problem \emph{per se}.  According to one
controversial view\cite{CEPI:GMH}, probabilities in any physical
theory need only be defined, and need only satisfy sum rules, to a
very good approximation, so that approximately consistent sets are all
that is ever needed.  Incorporating pragmatic observation into
fundamental theory in this way clearly, at the very least, raises
awkward questions.  Fortunately, it seems unnecessary.  There are good
reasons to expect\cite{Dowker:Kent:approach} to find exactly
consistent sets very close to a generic approximately consistent set,
so that even if only exactly consistent sets are permitted the
standard quasiclassical description can be recovered.  Note, though,
that none of the relevant exactly consistent sets will generally be
defined by Schmidt projections.

It could be argued that physically reasonable set selection criteria
should make predictions which vary continuously with structural
perturbations and perturbations in the initial conditions, and that
the instability of exact consistency under perturbation means that the
most useful consistency criteria are very likely to be approximate.
Certainly, there seems no reason in principle why a precisely defined
selection algorithm which gives physically sensible answers should be
rejected if it fails to exactly respect the consistency criterion.
For, once a single set has been selected, there seems no fundamental
problem in taking the decoherence functional probability weights to
represent precisely the probabilities of its fine-grained histories
and the probability sum rules to \emph{define} the probabilities of
coarse-grained histories.  On the other hand, allowing approximate
consistency raises new difficulties in identifying a single natural
set selection algorithm, since any such algorithm would have --- at
least indirectly --- to specify the degree of approximation tolerated.

These arguments over fundamentals, though, go beyond our scope here.
Our aim below is to investigate selection rules which might give
physically interesting descriptions of quantum systems, whether or not
they produce exactly consistent sets.  As we will see, it seems
surprisingly hard to find good selection rules even when we follow the
standard procedure in the decoherence literature and allow some degree
of approximate decoherence.

Mathematical definitions of approximate consistency were first
investigated by Dowker and Halliwell\cite{Dowker:Halliwell}, who
proposed a simple criterion --- the Dowker-Halliwell criterion, or DHC
--- according to which a set is approximately consistent to order
$\epsilon$ if the decoherence functional
\begin{equation} \label{decohfunct} 
 D_{\alpha \beta } = \langle\psi| C_{\beta}^\dagger C_{\alpha}
 |\psi\rangle
\end{equation} 
satisfies the equation
\begin{eqnarray}\label{DHC}
  |D_{\alpha\beta}| & \leq & \epsilon \,
    (D_{\alpha\alpha}D_{\beta\beta})^{1/2}, \quad \forall\,
    \alpha\neq\beta.
\end{eqnarray}
Approximate consistency criteria were analysed further in
ref. \cite{McElwaine:1}.  As refs. \cite{Dowker:Halliwell,McElwaine:1}
explain, the DHC has natural physical properties and is well adapted
for mathematical analyses of consistency.  We adopt it here, and refer
to the largest term,
\begin{equation} \label{dhp} 
\mbox{max} \{\,
  |D_{\alpha\beta}|(D_{\alpha\alpha}D_{\beta\beta})^{-1/2} \, : \,
  \alpha , \beta \in S \, , \alpha \neq \beta \, , \mbox{~and~}
  D_{\alpha\alpha} , D_{\beta\beta} \neq 0 \,\} \, ,
\end{equation} 
of a (possibly incomplete) set of histories $S$ as the
Dowker-Halliwell parameter, or DHP.

\label{nontrivpage}
A \emph{trivial} history $\alpha$ is one whose probability is zero,
$C_\alpha |\psi\rangle = 0$.  Many of the algorithms we discuss
involve, as well as the DHP, a parameter which characterises the
degree to which histories approach triviality.  The simplest
non-triviality criterion would be to require that all history
probabilities must be greater than some parameter $\delta$, i.e.\ that
\begin{equation}\label{absnontriv} 
D_{\alpha\alpha} > \delta \quad \mbox{for all histories $\alpha$.}
\end{equation} 
As a condition on a particular extension $\{P_i : i = 1,2 , \ldots \}$
of the history $\alpha$ this would imply that $\|P_iC_\alpha
|\psi\rangle\|^2 > \delta$ for all $i$.  This, of course, is an
absolute condition, which depends on the probability of the original
history $\alpha$ rather than on the relative probabilities of the
extensions and which implies that once a history with probability less
than $2\delta$ has been selected any further extension is forbidden.

It seems to us more natural to use criteria, such as the DHC, which
involve only relative probabilities.  It is certainly simpler in
practice: applying absolute criteria strictly would require us to
compute from first cosmological principles the probability to date of
the history in which we find ourselves.  We therefore propose the
following relative non-triviality criterion: an extension $\{P_i : i =
1,2 , \ldots \}$ of the non-trivial history $\alpha$ is non-trivial to
order $\delta$, for any $\delta$ with $0 < \delta < 1$, if
\begin{equation} \label{relnontriv} 
\|P_i \, C_\alpha |\psi\rangle\|^2 \geq \delta \| C_\alpha
|\psi\rangle\|^2 \quad \mbox{for all $i$.}
\end{equation}
We say that a set of histories $S$, which may be branch-dependent, is
non-trivial to order $\delta$ if every set of projections, considered
as an extension of the histories up to the time at which it is
applied, is non-trivial to order $\delta$.  In both cases we refer to
$\delta$ as the non-triviality parameter, or NTP.

An obvious disadvantage of applying an absolute non-triviality
criterion to branch-independent consistent sets is that, if the set
contains one history of probability less than or equal to $2 \delta$,
no further extensions are permitted.

Once again, though, our approach is pragmatic, and in order to cover
all the obvious possibilities we investigate below absolute
consistency and non-triviality criteria as well as relative ones.

\section{Repeated projections and consistency}\label{sec:repeated} 

One of the problems which arises in trying to define physically
interesting set selection algorithms is the need to find a way either
of preventing near-instantaneous repetitions of similar projections or
of ensuring that such repetitions, when permitted, do not prevent the
algorithm from making physically interesting projections at later
times.  It is useful, in analysing the behaviour of repeated
projections, to introduce a version of the DHC which applies to the
coincident time limit of sets of histories defined by smoothly
time-dependent projective decompositions.

To define this criterion, fix a particular time $t_0$, and consider
class operators $C_\alpha$ consisting of projections at times ${\bf t}
= (t_1, \ldots, t_n)$, where $t_n > t_{n-1} > \ldots > t_1 >
t_0$. Define the \emph{normalised histories} by
\begin{equation}\label{normhist}
  |\hat \alpha\rangle = \lim_{{\bf t'} \to {\bf t}} \frac{C_\alpha
      ({\bf t}') |\psi\rangle}{\|C_\alpha ({\bf t}') |\psi\rangle\|},
\end{equation}
where the limits are taken in the order $t'_1 \to t_1$ then $t'_2 \to
t_2$ and so on, whenever these limits exist.  Define the limit DHC
between two normalised histories $|\hat \alpha \rangle$ and $| \hat
\beta \rangle$ as
\begin{equation}\label{limitDHC}
  \langle \hat \alpha | \hat \beta\rangle \leq \epsilon \, .
\end{equation}
This, of course, is equivalent to the limit of the ordinary DHC when
the limiting histories exist and are not null.  It defines a stronger
condition when the limiting histories exist and at least one of them
is null, since in this case the limit of the DHC is automatically
satisfied.

If a set of histories is defined by a smoothly time-dependent
projective decomposition applied at two nearby times, it will contain
many nearly null histories, since $P_m P_n = 0$ for all $n \neq m$.
Clearly, in the limit as the time separation tends to zero, these
histories become null, so that the limit of the ordinary DHC is
automatically satisfied.  When do the normalised histories satisfy the
stronger criterion (\ref{limitDHC})?

Let $P(t)$ be a projection operator with a Taylor series at $t=0$,
\begin{equation}\label{taylorproj}
  P(t) = P + t \dot{P} + \frac{1}{2}t^2\ddot{P} + O(t^3) \, ,
\end{equation}
where $P=P(0)$, $\dot{P} = dP(t)/dt|_{t=0}$ and $\ddot{P} =
d^2P(t)/dt^2|_{t=0}$. Since $P^2(t) = P(t)$ for all $t$
\begin{equation} 
\begin{array}{rcl} 
  P + t \dot{P} + \frac{1}{2}t^2\ddot{P} + O(t^3) &=& [P + t \dot{P} +
  \frac{1}{2}t^2\ddot{P} + O(t^3)] [P + t \dot{P} +
  \frac{1}{2}t^2\ddot{P} + O(t^3)] \\ &=& P + t(P\dot{P} + \dot{P}P) +
  \frac{1}{2}t^2(P\ddot{P} + \ddot{P}P + 2\dot{P}^2) + O(t^3) \, .
\end{array}
\end{equation} 
This implies that
\begin{equation} \label{dPidenta}
  \dot{P} = P\dot{P} + \dot{P}P \, ,
\end{equation}
and
\begin{equation} \label{dPidentb}
\frac{1}{2}\ddot{P} = \frac{1}{2}P\ddot{P} + \frac{1}{2}\ddot{P}P +
\dot{P}^2 \,.
\end{equation}
Now consider a projective decomposition $\{P_k\}$ and the matrix
element
\begin{equation}
  \langle \psi | P_{m} P_{k} (t) P_{n} | \psi \rangle = \langle \psi |
  P_{m} P_{k} P_{n} | \psi \rangle + t \langle \psi | P_{m} \dot P_{k}
  P_{n} | \psi \rangle + \frac{1}{2} t^2 \langle \psi | P_{m} \ddot
  P_{k} P_{n} | \psi \rangle + O(t^3)\, . \label{matrixel1}
\end{equation}
Now $P_{m} P_{k} P_{n} = P_k \delta_{km} \delta_{kn}$, since the
projections are orthogonal, and
\begin{equation} 
\begin{array}{rcl} 
  P_{m} \dot P_{k} P_{n} &=& \delta_{km} (1-\delta_{kn}) \dot P_{k}
  P_{n} + \delta_{kn}(1-\delta_{km}) P_{m} \dot P_{k} \\ &=&
  \delta_{km} \dot P_{k} P_{n} + \delta_{kn} P_{m} \dot P_{k} -
  \delta_{km} \delta_{kn} \dot P_{k} \, ,
\end{array}
\end{equation} 
since $\dot P_{k} P_n = P_k \dot P_{k} P_n $ if $ k \neq n$ and $\dot
P_{k} P_k = (1 - P_k ) \dot P_{k} $.  (No summation convention applies
throughout this paper.)  From eq.~(\ref{dPidentb}) we have that
\begin{eqnarray}
  \frac{1}{2} P_{m} \ddot P_{k} P_{n} &=& \frac{1}{2} (\delta_{mk} +
  \delta_{nk}) P_{m} \ddot P_{k} P_{n} + P_{m} \dot P_{k}^2 P_{n} \, .
\end{eqnarray} 
Eq.~(\ref{matrixel1}) can now be simplified. To leading order in $t$
it is
\begin{eqnarray} 
  && \langle \psi | P_k | \psi \rangle + O(t) \qquad \mbox{if
    $k=m=n$,} \\&& t\langle \psi|\dot P_k P_n |\psi \rangle +O(t^2)
    \qquad \mbox{if $k=m$, $k\neq n$,} \\&& t\langle \psi| P_m \dot
    P_k |\psi \rangle +O(t^2) \qquad \mbox{if $k\neq m$, $k=n$, and}
    \\ \label{repproj} && t^2 \langle \psi | P_{m} \dot P_{k}^2 P_{n}
    | \psi \rangle + O(t^3) \qquad \mbox{if $k\neq m$, $k \neq n$.}
\end{eqnarray}

Now consider a smoothly time-dependent projective decomposition,
$\sigma(t) = \{ P(t) , \overline{ P } (t) \}$, defined by a
time-dependent projection operator and its complement.  Write $P=
P(0)$, and consider a state $|\phi\rangle$ such that $P|\phi\rangle
\neq 0 $ and $\overline{P}|\phi\rangle \neq 0 $.  We consider a set of
histories with initial projections $P, \overline{P}$, so that the
normalised history states at $t=0$ are
\begin{equation}
  \left\{ \frac{P|\phi\rangle}{\|P|\phi\rangle\|},
      \frac{\overline{P}|\phi\rangle}{\|\overline{P}|\phi\rangle\|}
      \right\} \, ,
\end{equation}
and consider an extended branch-dependent set defined by applying
$\sigma (t)$ on one of the branches --- say, the first --- at a later
time $t$.

The new normalised history states are
\begin{equation}\label{a1}
  \left\{ \frac{P(t)P|\phi\rangle}{\|P(t)P|\phi\rangle\|},
   \frac{\overline{P}(t)P|\phi\rangle}{\|\overline{P}(t)P|\phi\rangle\|},
   \frac{\overline{P}|\phi\rangle}{\|\overline{P}|\phi\rangle\|}
   \right\}.
\end{equation}
We assume now that $\dot{P}P|\phi\rangle \neq 0$, so that the limit of
these states as $t \to 0$ exists.  We have that
\begin{equation}
  \lim_{t\to0} \frac{(\overline P - t \dot{P}) P |\phi\rangle}{(t^2
    \langle \phi | P \dot P^2 P | \phi \rangle)^{1/2}} =
    \frac{-\dot{P} P |\phi\rangle }{\|\dot{P} P |\phi\rangle \|} \, ,
\end{equation}
so that the limits of the normalised histories are
\begin{equation}
  \left\{ \frac{P|\phi\rangle}{\|P|\phi\rangle\|},
   \frac{-\dot{P}P|\phi\rangle}{\|\dot{P}P|\phi\rangle\|},
   \frac{\overline{P}|\phi\rangle}{\|\overline{P}|\phi\rangle\|}
   \right\}.
\end{equation}
The only possibly non-zero terms in the limit DHC are
\begin{equation} \label{dblprojDHCterm}
  -\frac{\langle\phi| \overline{P} \dot{P} P |\phi\rangle }{\|
    \overline{P} |\phi\rangle \| \, \| \dot{P} P |\phi\rangle \|} =
    -\frac{\langle\phi| \overline{P} \dot{P} |\phi\rangle }{\|
    \overline{P} |\phi\rangle \| \, \| \overline{P} \dot{P}
    |\phi\rangle \|} \, ,
\end{equation}
which generically do not vanish.

Consider instead extending the second branch using $P(t)$ again. This
gives the set
\begin{equation}
  \left\{\frac{P|\phi\rangle}{\|P|\phi\rangle\|},
    \frac{\overline{P}|\phi\rangle}{\|\overline{P}|\phi\rangle\|},
    \frac{-P(t)\dot{P}P|\phi\rangle}{\|P(t)\dot{P}P|\phi\rangle\|},
    \frac{-\overline{P}(t)\dot{P}P|\phi\rangle}{\|
    \overline{P}(t)\dot{P}P |\phi\rangle\|} \right\}.
\end{equation}
Since $P\dot{P}P = 0$ the limit $t \to 0 $ exists and is
\begin{equation}
  \left\{\frac{P|\phi\rangle}{\|P|\phi\rangle\|},
    \frac{\overline{P}|\phi\rangle}{\|\overline{P}|\phi\rangle\|},
    \frac{-\dot{P}^2P|\phi\rangle}{\|\dot{P}^2P|\phi\rangle\|},
    \frac{-\dot{P}P|\phi\rangle}{\|\dot{P}P|\phi\rangle\|} \right\}.
\end{equation}
The DHC term between the first and third histories is
\begin{equation}\label{tripleprojDHCterm}
  -\frac{\langle\phi| P\dot{P}^2P |\phi\rangle}{\|P|\phi\rangle\| \,
    \|\dot{P}^2P |\phi\rangle\|} =
    -\frac{\|\dot{P}P|\phi\rangle\|^2}{\| P|\phi\rangle\| \,
    \|\dot{P}^2P |\phi\rangle\|} \, .
\end{equation}
This is always non-zero since $P\dot{P}|\phi\rangle \neq 0$.

For the same reason, extending the first branch again, or the third
branch, violates the limit DHC\@.  Hence, if projections are 
taken from a continuously parameterised set, and the limit DHC is used,
multiple re-projections will generically be forbidden.

The assumption that $\dot P P |\phi\rangle \neq 0$ can be relaxed. It
is sufficient, for example, that there is some $k$ such that
$\|P^{(j)}\|=0$ for all $j<k$ and that $P^{(k)} P |\phi\rangle \neq
0$, where $P^{(j)} = d^jP(t)/dt^j|_{t=0}$.

Note, finally, that it is easy to construct examples in which a single
re-projection is consistent.  For instance, let
\begin{equation}
  P = \left(
    \begin{array}{cc} I_{d_1} & 0 \\ 0 & 0 \end{array}
  \right) \quad \overline{P} = \left(
    \begin{array}{cc} 0 & 0 \\ 0 & I_{d_2} \end{array} 
  \right) \quad \dot{P} = \left(
    \begin{array}{cc} 0 & A^\dagger \\ A & 0 \end{array} \right)
  \quad |\phi\rangle = \left(
    \begin{array}{c} \sqrt{q} \, {\bf x}\\\sqrt{1-q} \, {\bf y},
    \end{array} \right)
\end{equation}
where ${\bf x}$ is a unit vector in $C^{d_1}$, ${\bf y}$ a unit vector
in $C^{d_2}$ and $A$ a $d_2 \times d_1$ complex matrix.
$\|P|\phi\rangle\| \neq 0,1$ implies that $q \neq 0,1$ and
$\dot{P}P|\phi\rangle \neq 0$ implies that $A{\bf x} \neq 0$. So from
eq. (\ref{dblprojDHCterm}) the DHC term is
\begin{equation}\label{tripleeg}
  -\frac{{\bf y}^\dagger A {\bf x}}{\|A {\bf x}\|} \, .
\end{equation}
If $d_2 \geq 2$ then ${\bf y}$ can be chosen orthogonal to $A {\bf x}$
and then eq.~(\ref{tripleeg}) is zero. The triple projection term
however, eq.~(\ref{tripleprojDHCterm}) is
\begin{equation}
  -\frac{\| A {\bf x} \|^2}{\| A^2{\bf x} \|} \, ,
\end{equation}
which is never equal to $0$ since $A{\bf x} \neq 0$.

\section{Schmidt projection algorithms}\label{sec:schmidt:alg} 

We turn now to the problem of defining a physically sensible set
selection algorithm which uses Schmidt projections, starting in this
section with an abstract discussion of the properties of Schmidt
projection algorithms.

We consider here dynamically generated algorithms in which initial
projections are specified at $t=0$, and the selected consistent set is
then built up by selecting later projective decompositions, whose
projections are sums of the Schmidt projection operators, as soon as
specified criteria are satisfied.  The projections selected up to time
$t$ thus depend only on the evolution of the system up to that time.
We will generally consider selection algorithms for branch-independent
sets and add comments on related branch-dependent selection
algorithms.

We assume that there is a set of Heisenberg picture Schmidt projection
operators $\{P_n (t)\}$ with continuous time dependence, defined even
at points where the Schmidt probability weights are degenerate, write
$P_n$ for $P_n (0)$, and let $I$ be the index set for projections
which do not annihilate the initial state, $I = \{ n : P_n
|\psi\rangle \neq 0 \}$.

We consider first a simple algorithm, in which the initial projections
are fixed to be the $P_n$ for $ n \in I$ together with their
complement $( 1 - \sum_n {P_n})$, and which then selects
decompositions built from Schmidt projections at the earliest possible
time, provided they are consistent.  More precisely, suppose that the
algorithm has selected a consistent set $S_k$ of projective
decompositions at times $t_0 , t_1 , \ldots , t_k$.  It then selects
the earliest time $t_{k+1} > t_k$ such that there is at least one
consistent extension of the set $S_k$ by a projective decomposition
formed from sums of Schmidt projections at time $t_{k+1}$.  In generic
physical situations, we expect this decomposition to be unique.
However, if more than one such decomposition exists, the one with the
largest number of projections is selected; if more than one
decomposition has the maximal number of projections, one is randomly
selected.

Though the limit DHC~(\ref{limitDHC}) can prevent trivial projections,
it does not generically do so here. The limit DHC terms between
histories $m$ and $n$ for an extension involving $P_k$ ($k \not \in
I$) are
\begin{equation} \label{initialDHC} 
  \lim_{t \to 0} \frac{| \langle \psi | P_{m} P_{k} (t) P_{n} | \psi
    \rangle | }{\| P_{m} | \psi \rangle \| \, \| P_{k} (t) P_{n} |
    \psi \rangle\|} = t \frac{| \langle \psi | P_{m} \dot P_{k}^2
    P_{n} | \psi \rangle | }{\| P_{m} | \psi \rangle \| \, \| \dot
    P_{k} P_{n} | \psi \rangle \|} = 0,
\end{equation} 
whenever $\| P_{m} | \psi \rangle \| $ and $\| \dot P_{k} P_{n} | \psi
\rangle\|$ are both non zero.  The first is non-zero by assumption;
the second is generically non-zero.  Thus the extension of all
histories by the projections $P_k \, (k \not \in I)$ and $\sum_{n \in
I} P_n$ satisfies the limit DHC.

Hence, if the initial projections do not involve all the Schmidt
projections, and if the algorithm tolerates any degree of approximate
consistency, whether relative or exact, then the DHC fails to prevent
further projections arbitrarily soon after $t=0$, introducing
histories with probabilities arbitrarily close to zero.
Alternatively, if the algorithm treats such projections by a limiting
process, then generically all the Schmidt projections at $t=0$ are
applied, producing histories of zero probability.  Similar problems
would generally arise with repeated projections at later times, if
later projections occur at all.

There would be no compelling reason to reject an algorithm which
generates unexpected histories of arbitrarily small or zero
probability, so long as physically sensible histories, of total
probability close to one, are also generated.  However, as we note in
subsection~\ref{sub:null} below and will see later in the analysis of
a physical example, this is hard to arrange.  We therefore also
consider below several ways in which small probability histories might
be prevented:
\begin{enumerate}
\item{} The initial state could be chosen so that it does not
precisely lie in the null space of any Schmidt projection. (See
subsection~\ref{sub:state}.)
\item{} An initial set of projections could somehow be chosen,
independent of the Schmidt projections, and with the property that for
every Schmidt projection at time zero there is at least one initial
history not in its null space.  (See subsection~\ref{sub:histories}.)
\item{} The algorithm could forbid zero probability histories by fiat
and require that the selected projective decompositions form an
exactly consistent set.  It could then prevent small probability
histories from occurring by excluding any projective decomposition
$\sigma (t)$ from the selected set if $\sigma (t)$ belongs to a
continuous family of decompositions, defined on some semi-open
interval $ ( t - \epsilon , t ]$, which satisfy the other selection
criteria. (See subsection~\ref{sub:adrian}.)
\item{} A parametrised non-triviality criterion could be used.  (See
subsection~\ref{sub:nontriv}.)
\item{} Some combination of parametrised criteria for approximate
consistency and non-triviality could be used. (See
subsection~\ref{sub:approxcon}.)
\end{enumerate}

We will see though, in this section and the next, that each of these
possibilities leads to difficulties.

\subsection{Choice of initial state}\label{sub:state}

In the usual description of experimental situations, ${\cal{H}}_1$
describes the system degrees of freedom, ${\cal{H}}_2$ those of the
apparatus (and/or an environment), and the initial state is a pure
Schmidt state of the form $ |\psi\rangle = |\psi_1 \rangle_1 \otimes
|\psi_2 \rangle_2$.  According to this description, probabilistic
events occur only after the entanglement of system and apparatus by
the measurement interaction.  It could, however, be argued that, since
states can never be prepared exactly, we can never ensure that the
system and apparatus are precisely uncorrelated, and the initial state
is more accurately represented by $ |\psi\rangle = |\psi_1 \rangle_1
\otimes |\psi_ 2\rangle_2 + \gamma |\phi\rangle$, where $\gamma$ is
small and $|\phi\rangle$ is a vector in the total Hilbert space chosen
randomly subject to the constraint that $ \langle \psi | \psi \rangle
= 1$.  A complete set of Schmidt projections $\{ P_n \}$, with $P_n
|\psi\rangle \neq 0 $ for all $n$, is then generically defined at
$t=0$, and any Schmidt projection algorithm which begins by selecting
all initial Schmidt projections of non-zero probability will include
all of the $ P_n $.

An obvious problem here, if relative criteria for approximate
consistency and non-triviality are used to identify subsequent
projections, is that the small probability initial histories constrain
the later projections just as much as the large probability history
which corresponds, approximately, to the Schmidt state
$|\psi_1\rangle_1 \otimes |\psi_2\rangle_2$ and which is supposed to
reproduce standard physical descriptions of the course of the
subsequent experiment.  If a branch-dependent selection algorithm is
used, a relative non-triviality criterion will not cause the small
probability initial histories to constrain the projections selected
later on the large probability branch, but a relative approximate
consistency criterion still will.

There seems no reason to expect the projections which reproduce
standard descriptions to be approximately consistent extensions of the
set defined by the initial Schmidt projections, and hence no reason to
expect to recover standard physics from a Schmidt projection
algorithm.  When we consider a simple model of a measurement
interaction in the next section we will see that, indeed, the initial
projections fail to extend to a physically natural consistent set.

If absolute criteria are used, on the other hand, we would expect
either that essentially the same problem arises, or that the small
probability histories do not constrain the projections subsequently
allowed and hence in particular do not solve the problems associated
with repeated projections, depending whether the probability of the
unphysical histories is large or small relative to the parameters
$\delta$ and $\epsilon^2$.

\subsection{Including null histories}\label{sub:null} 

If the initial state is Schmidt pure, or more generally does not
define a maximal rank Schmidt decomposition, a full set of Schmidt
projections can nonetheless generically be defined at $t=0$ --- which
we take to be the start of the interaction --- by taking the limit of
the Schmidt projections as $t \rightarrow 0^+$.  The normalised
histories corresponding to the projections of zero probability weight
can then be defined as above, if the relevant limits exist, and used
to constrain the subsequent projections in any algorithm involving
relative criteria.  Again, though, there seems no reason to expect
these constraints to be consistent with standard physical
descriptions.
 
\subsection{Redefining the initial conditions}\label{sub:histories}

The projections selected at $t=0$ could, of course, be selected using
quite different principles from those used in the selection of later
projections.  By choosing initial projections which are not
consistently extended by any of the decompositions defined by Schmidt
projections at times near $t=0$, we can certainly prevent any
immediate reprojection occurring in Schmidt selection algorithms.  We
know of no compelling theoretical argument against incorporating
projections into the initial conditions, but have found no natural
combination of initial projections and a Schmidt projection selection
algorithm that generally selects physically interesting sets.
  
\subsection{Exact consistency and a non-triviality criterion}
\label{sub:adrian}

Since many of the problems above arise from immediate reprojections,
it seems natural to look at rules which prevent zero probability
histories.  The simplest possibility is to impose precisely this
constraint, together with exact consistency and the rules that (i)
only one decomposition can be selected at any given time and (ii) no
projective decomposition can be selected at time $t$ if it belongs to
a continuous family of projections $\sigma (t)$, whose members would,
but for this rule, be selected at times lying in some interval $( t -
\epsilon , t ]$.  This last condition means that the projections
selected at $t=0$ are precisely those initially chosen and that no
further projections occur in the neighbourhood of $t=0$.
Unfortunately, as the model studied later illustrates, it also
generally prevents physically sensible projective decompositions being
selected at later times.  If it is abandoned, however, and if the
initial state $\ket{\psi}$ is a pure Schmidt state, then further
projections will be selected as soon as the interaction begins: in
other words, at times arbitrarily close to $t=0$.  Again, these
projections are generally inconsistent with later physically natural
projections.  On the other hand, if $\ket{\psi}$ is Schmidt-impure,
this is generally true of the initial projections.

All of these problems also arise in the case of branch-dependent set
selection algorithms.

\subsection{Exact consistency and a parametrised 
non-triviality criterion} \label{sub:nontriv}

Another apparently natural possibility is to require exact consistency
together with one of the parametrised non-triviality criteria
(\ref{absnontriv}) or (\ref{relnontriv}), rather than simply
forbidding zero probability histories.  A priori, there seem no
obvious problems with this proposal but, again, we will see that it
gives unphysical answers in the model analysed below, whether
branch-dependent or branch-independent selection algorithms are
considered.

\subsection{Approximate consistency and a parametrised 
non-triviality criterion}\label{sub:approxcon}

There are plausible reasons, apart from the difficulties of other
proposals, for studying algorithms which use approximate consistency
and parametrised non-triviality.  The following comments apply to both
branch-dependent and branch-independent algorithms of this type.

Physically interesting sets of projective decompositions --- for
example, those characterising the pointer states of an apparatus after
each of a sequence of measurements --- certainly form a set which is
consistent to a very good approximation.  Equally, in most cases
successive physically interesting decompositions define non-trivial
extensions of the set defined by the previous decompositions: if the
probability of a measurement outcome is essentially zero then, it
might plausibly be argued, it is not essential to include the outcome
in the description of the history of the system.  Moreover, a finite
non-triviality parameter $\delta$ ensures that, after a Schmidt
projective decomposition is selected at time $t$, there is a finite
time interval $[ t , t + \Delta t ]$ before a second decomposition can
be chosen.  One might hope that, if the parameters are well chosen,
the Schmidt projective decompositions at the end of and after that
interval will no longer define an approximately consistent extension
unless and until they correspond to what would usually be considered
as the result of a measurement-type interaction occurring after time
$t$.  While, on this view, the parameters $\epsilon$ and $\delta$ are
artificial, one might also hope that they might be eliminated by
letting them tend to zero in a suitable limit.

However, as we have already mentioned, in realistic physical
situations we should not necessarily expect any sequence of Schmidt
projective decompositions to define an exactly consistent set of
histories.  When the Schmidt projections correspond, say, to pointer
states, the off-diagonal terms of their decoherence matrix typically
decay exponentially, vanishing altogether only in the limit of
infinite time
separation\cite{GM:Hartle:classical,Caldeira:Leggett,Joos:Zeh,%
Dowker:Halliwell,Pohle,Paz:brownian:motion,%
Zurek:preferred:observables,Anastopoulos:Halliwell,%
Tegmark:Shapiro,Paz:Zurek:classicality,Hartle:spacetime,%
Zurek:transition}.  An algorithm which insists on exact consistency,
applied to such situations, will fail to select any projective
decompositions beyond those initially selected at $t=0$ and so will
give no historical description of the physics.  We therefore seem
forced, if we want to specify a Schmidt projection set selection
algorithm mathematically, to introduce a parameter $\epsilon$ and to
accept sets which are approximately consistent to order $\epsilon$ and
then, in the light of the preceding discussion, to introduce a
non-triviality parameter $\delta$ in order to try to prevent
unphysical projective decompositions being selected shortly after
$t=0$.  This suggests, too, that the best that could be expected in
practice from an algorithm which uses a limit in which $\epsilon$ and
$\delta$ tend to zero is that the resulting set of histories describes
a series of events whose time separations tend to infinity.

A parameter-dependent set selection algorithm, of course, leaves the
problem of which values the parameters should take.  One might hope,
at least, that there is a range of values for $\epsilon$ and $\delta$
over which the selected set varies continuously and has essentially
the same physical interpretation.  An immediate problem here is that,
if the first projective decomposition selected after $t=0$ defines a
history which only just satisfies the non-triviality condition, the
decomposition will, once again, have no natural physical
interpretation and will generally be inconsistent with the physically
natural decompositions which occur later.  We will see that, in the
simple model considered below, this problem cannot be avoided with an
absolute consistency criterion.
 
Suppose now that we impose the absolute non-triviality condition that
all history probabilities must be greater than $\delta$ together with
the relative approximate consistency criterion that the modulus of all
DHC terms is less than $\epsilon$.  The parameters $\epsilon$ and
$\delta$ must be chosen so that these projections stop being
approximately consistent before they become non-trivial otherwise
projections will be made as soon as they produce histories of
probability exactly $\delta$, in which case the non-triviality
parameter, far from eliminating unphysical histories, would be
responsible for introducing them.

Let $t_\epsilon$ denote the latest time that the extension with
projection $P_k(t)$ is approximately consistent and $t_\delta$ the
earliest time at which the extension is nontrivial. We 
see from~(\ref{initialDHC}) that, to lowest order in $t$,
\begin{eqnarray}
  t_\delta &=& \sqrt{\delta} \|\dot P_{k} P_{n} | \psi \rangle \|^{-1}
  \\ t_\epsilon &=& \epsilon \frac{\| P_{m} | \psi \rangle \| \, \|
  \dot P_{k} P_{n} | \psi \rangle \|}{|\langle \psi | P_{m} \dot
  P_{k}^2 P_{n} | \psi \rangle|}.
\end{eqnarray}
$t_\delta > t_\epsilon$ implies
\begin{equation}\label{edinequality}
  \sqrt{\delta} |\langle \psi | P_{m} \dot P_{k}^2 P_{n} | \psi
  \rangle| > \epsilon \| P_{m} | \psi \rangle \|\, \| \dot P_{k} P_{n}
  | \psi \rangle \|^2.
\end{equation}
Thus we require $\delta > \epsilon^2$, up to model-dependent numerical
factors: this, of course, still holds if we use a relative
non-triviality criterion rather than an absolute one.

This gives, at least, a range of parameters in which to search for
physically sensible consistent sets, and over which there are natural
limits --- for example $\mbox{lim}_{\delta \rightarrow 0}
\mbox{lim}_{\epsilon \rightarrow 0}$.  We have, however, as yet only
looked at some model-independent problems which arise in defining
suitable set selection rules.  In order to gain some insight into the
physical problems, we look next at a simple model of
system-environment interactions.

\section{A simple spin model}\label{sec:spin:intro} 

We now consider a simple model in which a single spin half particle,
the system, moves past a line of spin half particles, the environment,
and interacts with each in turn.  This can be understood as modelling
either a series of measurement interactions in the laboratory or a
particle propagating through space and interacting with its
environment.  In the first case the environment spin half particles
represent pointers for a series of measuring devices, and in the
second they could represent, for example, incoming photons interacting
with the particle.

Either way, the model omits features that would generally be
important.  For example, the interactions describe idealised sharp
measurements --- at best a good approximation to real measurement
interactions, which are always imperfect.  The environment is
represented initially by the product of $N$ particle states, which are
initially unentangled either with the system or each other.  The only
interactions subsequently considered are between the system and the
environment particles, and these interactions each take place in
finite time.  We assume too, for most of the following discussion,
that the interactions are distinct: the $k^{\mbox{\scriptsize th}}$ is
complete before the $(k+1)^{\mbox{\scriptsize th}}$ begins.  It is
useful, though, even in this highly idealised example, to see the
difficulties which arise in finding set selection algorithms: we take
the success of a set selection algorithm here to be a necessary, but
not sufficient, condition for it to be considered as a serious
candidate.

\subsection{Definition of the model} 

We use a vector notation for the system states, so that if ${\bf u}$
is a unit vector in $R^3$ the eigenstates of $\sigma. {\bf u }$ are
represented by $| \bf \pm u \rangle$.  With the pointer state analogy
in mind, we use the basis $\{ |\uparrow\rangle_k ,
|\downarrow\rangle_k \}$ to represent the $k^{\mbox{\scriptsize th}}$
environment particle state, together with the linear combinations
$|\pm\rangle_k = (|\uparrow\rangle_k \pm |\downarrow\rangle_k
)/\sqrt{2}$.  We compactify the notation by writing environment states
as single kets, so that for example $ |\uparrow\rangle_1 \otimes
\cdots \otimes |\uparrow\rangle_n $ is written as $| \uparrow_1 \ldots
\uparrow_n \rangle$, and we take the initial state $|\psi(0)\rangle$
to be $|{\bf v}\rangle \otimes | \uparrow_1 \ldots \uparrow_n
\rangle$.

The interaction between the system and the $k^{\mbox{\scriptsize th}}$
environment particle is chosen so that it corresponds to a measurement
of the system spin along the ${\bf u}_k$ direction, so that the states
evolve as follows:
\begin{eqnarray} \label{measurementinteraction}
  |{\bf u}_k\rangle \otimes |\uparrow\rangle_k & \to & |{\bf
    u}_k\rangle \otimes |\uparrow\rangle_k \, , \\ |{\bf -u}_k \rangle
    \otimes |\uparrow\rangle_k & \to & |{\bf -u}_k \rangle \otimes
    |\downarrow\rangle_k.
\end{eqnarray}
A simple unitary operator that generates this evolution is
\begin{equation}\label{Ukdef} 
  U_k( t ) = P({\bf u}_k) \otimes I_k + P({\bf -u}_k) \otimes
  \mbox{e}^{-i\theta_k(t) F_k} \, ,
\end{equation}
where $P({\bf x}) = |{\bf x}\rangle \langle{\bf x}|$ and $F_k =
i|\downarrow\rangle_k \langle\uparrow|_k - i|\uparrow\rangle_k
\langle\downarrow|_k$.  Here $\theta_k(t)$ is a function defined for
each particle $k$, which varies from $0$ to $\pi/2$ and represents how
far the interaction has progressed.  We define $P_k ({ \pm}) = |{ \pm
}\rangle_k \langle{ \pm}|_k $, so that $F_k = P_k (+)-P_k (-)$.

The Hamiltonian for this interaction is thus
\begin{equation}
  H_{k}(t) = i\dot U_k (t) U_k^\dagger (t) \\ = \dot \theta_k(t)
  P({\bf -u}_k) \otimes F_k \, ,
\end{equation}
in both the Schr\"odinger and Heisenberg pictures.  We write the
extension of $U_k$ to the total Hilbert space as
\begin{equation}\label{Vkdef} 
  V_k = P({\bf u}_k) \otimes I_1 \otimes \cdots \otimes I_n + P({\bf
    -u}_k) \otimes I_1 \otimes \cdots \otimes I_{k-1} \otimes
    \mbox{e}^{-i\theta_k(t) F_k} \otimes I_{k+1} \otimes \cdots
    \otimes I_n \,.
\end{equation}
We take the system particle to interact initially with particle $1$
and then with consecutively numbered ones, and there is no interaction
between environment particles, so that the evolution operator for the
complete system is
\begin{equation}
  U(t) = V_n(t) \ldots V_1(t) \, ,
\end{equation}
with each factor affecting only the Hilbert spaces of the system and
one of the environment spins.

We suppose, finally, that the interactions take place in disjoint time
intervals and that the first interaction begins at $t=0$, so that the
total Hamiltonian is simply
\begin{equation} 
 H (t ) = \sum_{k=1}^n H_k (t) \, ,
\end{equation} 
and we have that $\theta_1 (t) > 0 $ for $t > 0$ and that, if
$\theta_k(t) \in ( 0,\pi/2 ) $, then $\theta_i(t) = \pi/2
{\rm~for~all~} i < k$ and $\theta_i(t) = 0 {\rm~for~all~} i >k$.

\subsection{Classification of Schmidt projection consistent sets in
the model} 

For generic choices of the spin measurement directions,
in which no adjacent pair of the vectors $\{{\bf v},{\bf u}_1, \ldots
,{\bf u}_n\}$ is parallel or orthogonal, the exactly consistent
branch-dependent sets defined by the Schmidt projections onto the
system space can be completely classified in this model. The following
classification theorem is proved in ref.\cite{McElwaine:2}:

\vspace{.5\baselineskip}
\noindent\emph{Theorem}\qquad 
In the spin model defined above, suppose that no adjacent pair of the
vectors $\{{\bf v},{\bf u}_1, \ldots ,{\bf u}_n\}$ is parallel or
orthogonal.  Then the histories of the branch-dependent consistent
sets defined by Schmidt projections take one of the following forms:
\begin{description} 
\item[(i)] a series of Schmidt projections made at times between the
interactions --- i.e.\ at times $t$ such that $\theta_k (t) = {0
{\rm~or~} \pi/2} {\rm~for~all~} k$.
\item[(ii)] a series as in (i), made at times $t_1 , \ldots , t_n$,
together with one Schmidt projection made at any time $t$ during the
interaction immediately preceding the last projection time $t_n$.
\item[(iii)] a series as in (i), together with one Schmidt projection
made at any time $t$ during an interaction taking place after $t_n$.
\end{description} 
Conversely, any branch-dependent set, each of whose histories takes
one of the forms (i)-(iii), is consistent.  \vspace{.5\baselineskip}

\noindent We assume below that the set of spin measurement directions 
satisfies the condition of the theorem: since this can be ensured by an
arbitrarily small perturbation, this seems physically reasonable.  The
next sections explain, with the aid of this classification, the
results of various set selection algorithms applied to the model.

\section{Application of selection algorithms to the spin model}
\label{app}

We can define a natural consistent set which reproduces the standard
historical account of the physics of the separated interaction spin
model by selecting the Schmidt projections at all times between each
successive spin measurement.  A set of this type ought to be produced
by a good set selection algorithm, either as the selected set itself
or, perhaps, a subset.  The first three subsections below describe the
results actually produced by various set selection algorithms applied
to the spin model.  All of these algorithms are dynamical, in the
sense that the decision whether to select projections at time $t$, and
if so which, depends only on the evolution of the state vector up to
time $t$.  The following two subsections discuss how these results are
affected by altering the initial conditions of the model.  In the next
subsection we consider a selection algorithm which is quasi-dynamical,
in the sense that the decisions at time $t$ depend on the evolution of
the state vector up to and just beyond $t$.  We summarise our
conclusions in the last subsection.

\subsection{Exact limit DHC consistency} 

Since any projective decomposition at time $t$ defines an exactly
consistent set when there is only one history up to that time, a
Schmidt projection selection algorithm without a non-triviality
criterion will immediately make a projection.  The normalised
histories are defined as
\begin{equation} 
  \lim_{t \to 0} P_\pm(t) |\psi\rangle / \|P_\pm(t) |\psi\rangle \| \,
  ,
\end{equation} 
where $P_\pm(t)$ denotes the Schmidt projections at time $t$.  The
Schmidt states to first order in $\omega = \theta_1(t)$ are
\begin{equation}
 |{\bf v}\rangle \otimes |\uparrow_1\ldots\uparrow_n\rangle - i
  \omega/2 (1 - {\bf u}_1. {\bf v}) |{\bf v}\rangle \otimes
  |\downarrow_1\uparrow_2\ldots\uparrow_n\rangle
\end{equation}
and
\begin{equation}
 |{\bf u}_1 \wedge {\bf v}| |{\bf -v}\rangle \otimes
  |\downarrow_1\uparrow_2\ldots\uparrow_n\rangle + i \omega/2 \sqrt{
  \frac{ 1 - {\bf u}_1. {\bf v} }{ 1 + {\bf u}_1. {\bf v} } } |{\bf
  -v}\rangle \otimes |\uparrow_1\ldots\uparrow_n\rangle \, ,
\end{equation}
so the normalised histories are
\begin{equation}
  \{ |{\bf v}\rangle \otimes
  |\uparrow_1\uparrow_2\ldots\uparrow_n\rangle , |{\bf -v}\rangle
  \otimes |\downarrow_1\uparrow_2\ldots\uparrow_n\rangle \}\,.
\end{equation}
The limit DHC term for one projection at time $0$ and another during
interaction $k$ at time $t$ is
\begin{equation}
\begin{array}{ll}  \displaystyle
  \cos\phi & \mbox{for $k = 1$,} \\ \displaystyle \frac{\sin^2\phi\,
  |{\bf u}_1.  {\bf u}_2| |{\bf v} \wedge ( {\bf u}_1 \wedge {\bf u}_2
  )|}{N_2(\phi) [1 - ({\bf v}. {\bf u}_1)^2 N_2^2(\phi)]^{1/2}} &
  \mbox{for $k = 2$,} \\[2ex] \displaystyle \frac{ \lambda_{2 (k-1)}
  N_k (\phi) | {\bf v} \wedge ( {\bf u}_1 \wedge {\bf u}_2 )|}{ [1 -
  \lambda^2_{0(k-1)} N^2_k (\phi)]^{1/2}} & \mbox{for $k > 2$} \, ,
\end{array} 
\end{equation}
where $\phi = \theta_k(t)$.  Here we define
\begin{equation}
\lambda_{ij} = \prod_{k=i}^{j-1} |{\bf u}_k. {\bf u}_{k+1}| \,,
\end{equation} 
with the convention that $\lambda_{ij} = 1$ for $j \leq i$, and
\begin{equation} 
N_k ( \phi ) = | A_k ( \phi ) {\bf u}_{k-1} | \, ,
\end{equation}
where
\begin{equation}
A_k (\phi ) = P( {\bf u}_k ) + \cos \phi\, \overline P ( {\bf u}_k )
\, ,
\end{equation} 
where $P ( {\bf u}_k )$ is the projection onto the vector $ {\bf u}_k$
in $R^3$, and $\overline{P} ( {\bf u}_k )$ its complement.

Whether the algorithm is taken to be branch-dependent or
branch-independent, the only future Schmidt projections which are
consistent with the initial projections are thus those between the
first and second interactions, and the projections selected will be at
the end of the first interaction. The state at this time is
\begin{equation}
  |\psi(1)\rangle = |{\bf u_1} \rangle \langle{\bf u}_1 | {\bf
  v}\rangle \otimes |\uparrow_1\ldots \uparrow_n\rangle + |{\bf -u}_1
  \rangle \langle{\bf -u}_1 | {\bf v}\rangle \otimes
  |\downarrow_1\uparrow_2\ldots\uparrow_n\rangle,
\end{equation}
The time evolved histories are
\begin{eqnarray}
  |h_1(t)\rangle &=& |{\bf u_1} \rangle \langle{\bf u}_1 | {\bf
  v}\rangle \otimes |\uparrow_1 \ldots \uparrow_n\rangle + |{\bf -u}_1
  \rangle\langle{\bf -u}_1 | {\bf v}\rangle \otimes |\uparrow_1 \ldots
  \uparrow_n\rangle \\ |h_2(t)\rangle &=& |{\bf u_1} \rangle
  \langle{\bf u}_1 | {\bf -v}\rangle \otimes |\downarrow_1\uparrow_2
  \ldots \uparrow_n\rangle - |{\bf -u}_1 \rangle \langle{\bf -u}_1 |
  {\bf -v}\rangle \otimes |\downarrow_1\uparrow_2 \ldots
  \uparrow_n\rangle
\end{eqnarray}
so the new normalised histories are
\begin{eqnarray}
  \{ |{\bf u}_1\rangle \otimes |\uparrow_1\ldots \uparrow_n\rangle,
  |{\bf u}_1\rangle \otimes |\downarrow_1\uparrow_2 \ldots
  \uparrow_n\rangle,\\ |{\bf -u}_1\rangle \otimes
  |\uparrow_1\rangle\ldots\uparrow_n\rangle, |{\bf -u}_1\rangle
  \otimes |\downarrow_1\uparrow_2 \ldots \uparrow_n\rangle\}.
\end{eqnarray}
Since no future Schmidt projections are consistent with those
selected, the algorithm clearly fails to produce the correct set.

\subsection{Exact consistency and non-triviality} 

Suppose that, instead of using the limit DHC, we consider only sets
defined by decompositions at different times and require exact
consistency.  As explained earlier, without a non-triviality criterion
this leads to an ill-defined algorithm: the initial projections at
$t=0$ produce a null history, and the Schmidt projections at all times
greater than zero are consistent with these initial projections, so
that no minimal non-zero time is selected by the algorithm.

Introducing a non-triviality criterion removes this problem.  Suppose,
for example, we impose the absolute criterion $D_{\alpha\alpha} \geq
\delta$ for all histories $\alpha$.  Since any physically reasonable
$\delta$ would have to be extremely small, let us assume $\delta \ll |
{\bf u}_i \wedge {\bf u}_j |$.  The first projections after $t=0$ are
then selected at the first time when $D_{\alpha\alpha} = \delta$,
which occurs during the first interaction.  Whether or not
branch-dependent projections are allowed, the only other Schmidt
projections which can consistently be selected then take place at the
end of the first interaction, and it again follows from the
classification theorem that no further projections can take place.
Again, by making projections too early, this algorithm fails to
produce the correct consistent set.

A suitably large value of $\delta$ could ensure that no extension will
occur until later interactions but, generically, the first extension
made after $t=0$ will take place during an interaction rather than
between interactions, and the classification theorem ensures that no
more than four histories will ever be generated.

The same problems arise if the non-triviality criterion is taken to be
relative rather than absolute.  It is possible to do better by
fine-tuning the parameters: for example, if branch independent
histories are used, a relative non-triviality criterion is imposed and
$ \delta = (1 - |{\bf u}_k. {\bf u}_{k+1}|)/2$ for all $k =
0,\ldots,n-1$, then projections will occur at the end of each
interaction producing the desired set of histories.  This, though, is
clearly not a satisfactory procedure.

\subsection{Approximate consistency and non-triviality} 

One might wonder if these problems can be overcome by relaxing the
standards of consistency, since a projection at a very small time will
be approximately consistent --- according to absolute measures of
approximate consistency, at least --- with projections at the end of
the other interactions.  However, this approach too runs into
difficulties, whether relative or exact criteria are used.

Consider first a branch-dependent set selection algorithm which uses
the absolute non-triviality criterion $D_{\alpha\alpha} \geq \delta$
for all $\alpha$, and the absolute criterion for approximate
consistency $|D_{\alpha\beta}| \leq \epsilon$ for all $\alpha \neq
\beta$.  No history with probability less than $2\delta$ will thus be
extended, since if it were one of the resultant histories would have
probability less than $\delta$.

Any history $\alpha$ with a probability less than or equal to
$\epsilon^2$ will automatically be consistent with any history $\beta$
according to this criterion, since $|D_{\alpha\beta}| \leq
(D_{\alpha\alpha} D_{\beta\beta})^{1/2} \leq (\epsilon^2 \cdot
1)^{1/2} = \epsilon$.  Therefore if $\delta \leq \epsilon^2$ then
histories of probability $\delta$ will be consistent with all other
histories.  The first projection after $t=0$ will be made as soon as
the non-triviality criterion permits, when the largest Schmidt
eigenvalue is $1-\delta$.  Other projections onto the branch defined
by the largest probability history will follow similarly as the
Schmidt projections evolve.  The final set of histories after $n$
projections will thus consist of one history with probability
$1-n\delta$ and $n$ histories with probability $\delta$ --- clearly
far from the standard picture.

Suppose now that $\delta > \epsilon^2$.  The probabilities for
histories with projection in the first interval, at time $t$ with
$\theta_1 (t) = \omega$, are
\begin{equation} 
   1/2 [1 - \sqrt{1- \sin^2\omega |{\bf v} \wedge {\bf u}_1|^2}].
\end{equation}
The first projection will therefore be made when
\begin{equation} \label{tdef} 
\theta_1 (t) = \omega \simeq 2\sqrt{\delta} |{\bf v} \wedge {\bf
  u}_1|^{-1} \, ,
\end{equation} 
producing histories of probabilities $\delta$ and $(1- \delta )$.  The
next projections selected will necessarily extend the history of
probability $(1- \delta )$, since the absolute non-triviality
criterion forbids further extensions of the other history.  We look
first at projections taking place at a later time $t'$, with $\theta_1
(t') = \phi$, during the first interaction, and define $N_1 (\omega )
= ( 1 - \sin^2\omega |{\bf v} \wedge {\bf u}_1|^2 )^{1/2} $.  Of the
probabilities of the extended histories, the smaller is
\begin{eqnarray}\nonumber
  && 1/4 [ 1 + N_1(\omega) ] \bigg\{ 1 - N^{-1}_1(\omega)
  N^{-1}_1(\phi) [({\bf v} . {\bf u}_1)^2 + \cos\phi \cos\omega
  \cos(\phi-\omega) |{\bf v} \wedge {\bf u}_1 |^2 ] \bigg\} \\ &=& 1/4
  |{\bf v} \wedge {\bf u}_1|^2 (\omega-\phi)^2 [ 1 + O(\omega) +
  O(\phi)] \, ,
\end{eqnarray} 
Therefore this extension will be non-trivial when
\begin{equation}
  \phi \simeq \omega + 2\sqrt{\delta} |{\bf v} \wedge {\bf u}_1|^{-1}
= 4\sqrt{\delta} |{\bf v} \wedge {\bf u}_1|^{-1} + O(\delta).
\end{equation}
The largest off-diagonal element in the decoherence matrix for this
extension is
\begin{eqnarray} \label{soreproj}
  1/4 N_1^{-1}(\phi) |{\bf v} \wedge {\bf u}_1|^2 \cos\phi \sin\omega
  \sin(\phi-\omega) = \delta + O(\delta^3).
\end{eqnarray}
Unless $\delta > \epsilon$, then, this extension is selected together,
again, with a series of further extensions generating small
probability histories.

Suppose now that $\delta > \epsilon$.  The term on the left hand side
of eq.~(\ref{soreproj}) increases monotonically until $\phi \simeq
\pi/4$, and then decreases again as $\phi \rightarrow \pi/2$.  For
$\phi \simeq \pi/2$, it equals
\begin{eqnarray}
 1/2 \sqrt{\delta} \cos \phi |{\bf v} \wedge {\bf u}_1| |{\bf v}. {\bf
  u}_1|^{-1} [1 + O(\cos\phi)] \,.
\end{eqnarray}
Hence the approximate consistency criterion is next satisfied when
\begin{equation}
  \phi = \pi/2 - \frac{2\epsilon |{\bf v}. {\bf u}_1|}{\sqrt{\delta}
        |{\bf v} \wedge {\bf u}_1|} + O(\epsilon^2/\delta) \, ,
\end{equation}
and this extension is also non-trivial unless $ \bf v $ and $ {\bf
u}_1 $ are essentially parallel, which we assume not to be the case.
In this case, then, projections are made towards the beginning and
towards the end of the first interaction, and a physically reasonable
description of the first measurement emerges.

This description, however, cannot generally be consistently extended
to describe the later measurements.  If we consider the set of
histories defined by the Schmidt projections at time $t$, given by
eq.~(\ref{tdef}) above, together with the Schmidt projections at time
$t''$ such that $\theta_k (t'') = \phi$ for some $k>1$, we find that
the largest off-diagonal decoherence matrix element is
\begin{equation} \label{tsep} 
1/2 \sqrt{\delta} \lambda_{2(k-1)} N_k(\phi) |{\bf v} \wedge {\bf
  u}_1| | {\bf v} \wedge ( {\bf u}_1 \wedge {\bf u}_2)|[ 1 +
  O(\sqrt\omega)] \,.
\end{equation}

Since we have chosen $\epsilon < \delta$ to prevent multiple
projections, and since the other terms are not small for generic
choices of the vectors, the set generally fails to satisfy the
criterion for approximate consistency.  Note, however, that if all the
measurement directions are apart by an angle greater than equal to
some $\theta >0$, then $\lambda_{2(k-1)}$ decreases exponentially with
$k$.  After a large enough number (of order $O(-\log \epsilon)$) of
interactions have passed the algorithm will select a consistent
extension, and further consistent extensions will be selected at
similar intervals.  The algorithm does thus eventually produce
non-trivial consistent sets, though the sets produced do not vary
smoothly with $\epsilon$ and do not describe the outcome of most of
the spin measurements.

The reason this algorithm, and similar algorithms using approximate
consistency criteria, fail is easy to understand.  The off-diagonal
decoherence matrix component in a set defined by the Schmidt
projections at time $t$ together with Schmidt projections during later
interactions is proportional to $\sin \omega \cos \omega$, together
with terms which depend on the angles between the vectors.  The
decoherence matrix component for a set defined by the projections at
time $t$, together with Schmidt projections at a second time $t'$ soon
afterwards is proportional to $\sin^2(\phi-\omega)$.  The obstacle to
finding non-triviality and approximate consistency criteria that can
prevent reprojections in the first interaction period, yet allow
interactions in later interaction periods, is that when $(\phi -
\omega)$ is small the second term is generally smaller than the first.

Using a relative non-triviality criterion makes no difference, since
the branchings we consider are from a history of probability close to
$1$, and using the DHC instead of an absolute criterion for
approximate consistency only worsens the problem of consistency of
later projections, since the DHC alters eq.~(\ref{tsep}) by a factor
of $1/\sqrt{\delta}$, leaving a term which is generically of order
unity.  Requiring branch-independence, of course, only worsens the
problems.

\subsection{Non-zero initial Schmidt eigenvalues}

We now reconsider the possibility of altering the initial conditions
in the context of the spin model.  Suppose first that the initial
state is not Schmidt degenerate.  For example, as the initial
normalised histories are $\{ |{\bf v}\rangle \otimes
|\uparrow_1\ldots\uparrow_n\rangle, |{\bf -v}\rangle \otimes
|\downarrow_1\uparrow_2\ldots\uparrow_n\rangle\}$ a natural ansatz is
\begin{equation}
  |\psi(0)\rangle = \sqrt{p_1}|{\bf v}\rangle
  \otimes|\uparrow_1\ldots\uparrow_n\rangle + \sqrt{p_2} |{\bf
  -v}\rangle \otimes |\downarrow_1\uparrow_2\ldots\uparrow_n\rangle
  \,.
\end{equation}
Consider now a set of histories defined by Schmidt projections at
times $0$ and a time $t$ during the $k^{\mbox{\scriptsize th}}$
interaction for $k >2$, so that $\theta_1 (t) = \theta_2 (t) = \pi/2$.
The moduluses of the non-zero off-diagonal elements of the decoherence
matrix are
\begin{equation}
  1/2 \sqrt{p_1p_2} | {\bf v} \wedge [{\bf u}_1 \wedge {\bf u}_2 ] |
\lambda_{2k} \,.
\end{equation}
Generically, these off-diagonal elements are not small, so that the
perturbed initial conditions prevent later physically sensible
projections from being selected.

\subsection{Specifying initial projections}

We consider now the consequence of specifying initial projections in
the spin model. Suppose the initial projections are made using $P({\bf
\pm h}) \otimes I_E$. The modulus of the non-zero off-diagonal
elements of the decoherence matrix for a projection at time $t$ during
interaction $k$, for $k>2$, is
\begin{equation} 
 1/4 | {\bf h} \wedge {\bf v}|\, |{\bf h} \wedge {\bf u}_1 |
\lambda_{1(k-1)} N_k ( \theta_k (t) ) \, ,
\end{equation} 
and again we see that physically natural projections generically
violate the approximate consistency criterion.
 
It might be argued that the choice of initial projections given by
${\bf h} = {\bf \pm v}$ is particularly natural.  This produces an
initial projection on to the initial state, with the other history
undefined unless a limiting operation is specified.  If the limit of
the normalised histories for initial projections ${\bf h'} \rightarrow
{\bf h}$ is taken, the normalised histories are simply $|{\bf \pm
h}\rangle$. If an absolute consistency criterion is used the null
history will not affect future projections and the results will be the
same as if no initial projection had been made. If, on the other hand,
the limit DHC is used then the consistency criterion is the same as
for general ${\bf h}$, that is ${\bf h}$ must be parallel to ${\bf
u}_1$.  This requires that the initial conditions imposed at $t=0$
depend on the axis of the first measurement, and still fails to permit
a physically natural description of later measurements.

\subsection{A quasi-dynamical algorithm}\label{quasi} 

For completeness, we include here an algorithm which, though not
strictly dynamical, succeeds in selecting the natural consistent set
to describe the spin model.  In the spin model as defined, it can be
given branch-dependent or branch-independent form and selects the same
set in either case.  In the branch-independent version, the Schmidt
projections are selected at time $t$ provided that they define an
exactly consistent and non-trivial extension of the set defined by
previously selected projections \emph{and} that this extension can
itself be consistently and non-trivially extended by the Schmidt
projections at time $t+ \epsilon$ for every sufficiently small
$\epsilon > 0$.\footnote{Alternatively, a limiting condition can be
used.}  In the branch-dependent version, the second condition must
hold for at least one of the newly created branches of non-zero
probability in the extended set.

It follows immediately from the classification theorem that no Schmidt
projections can be selected during interactions, since no exactly
consistent set of Schmidt projections includes projections at two
different times during interactions.  The theorem also implies that
the Schmidt projections are selected at the end of each interval
between interactions, so that the selected set describes the outcomes
of each of the measurements.

\subsection{Comments}

The simple spin model used here illustrates the difficulty in encoding
our physical intuition algorithmically.  The model describes a number
of separated interactions, each of which can be thought of as a
measurement of the system spin.  There is a natural choice of
consistent set, given by the projections onto the system spin states
along the measured axes at all times between each of the
measurements.\footnote{Strictly speaking, there are many equivalent
consistent sets, all of which include the Schmidt projections at some
point in time between each measurement and at no time during
measurements, and all of which give essentially the same physical
picture.}  This set does indeed describe the physics of the system as
a series of measurement events and assigns the correct probabilities
to those events.  Moreover, the relevant projections are precisely the
Schmidt projections.

We considered first a series of Schmidt projection set selection
algorithms which are dynamical, in the sense that the projections
selected at time $t$ depend only on the physics up to that time.
Despite the simplifying features of the models, it seems very hard to
find a dynamical Schmidt projection set selection algorithm which
selects a physically natural consistent set and which is not
specifically adapted to the model in question.

It might be argued that the very simplicity of the model makes it an
unsuitable testing ground for set selection algorithms.  It is
certainly true that more realistic models would generally be expected
to allow fewer exactly consistent sets built from Schmidt projections:
it is not at all clear that any non-trivial exactly consistent sets of
this type should be expected in general.  However, we see no way in
which all the problems encountered in our discussion of dynamical set
selection algorithms can be evaded in physically realistic models.

We have, on the other hand, seen that a simple quasi-dynamical set
selection algorithm produces a satisfactory description of the spin
model.  However, as we explain in the next section, there is another
quite general objection which applies both to dynamical set selection
algorithms and to this quasi-dynamical algorithm.

\section{The problem of recoherence} \label{sec:recoherence} 

The set selection algorithms above rely on the decoherence of the
states of one subsystem through their interactions with another.  This
raises another question: what happens when decoherence is followed by
recoherence?

For example, consider a version of the spin model in which the system
particle initially interacts with a single environment particle as
before, and then re-encounters the particle, reversing the
interaction, so that the evolution takes the form
\begin{equation} \label{recoh} 
\begin{array}{rcl} 
a_1 | {\bf u}\rangle \otimes |\uparrow_1\rangle + a_2 |{\bf -u}
\rangle \otimes |\uparrow_1\rangle & \to & a_1 | {\bf u}\rangle
\otimes |\uparrow_1\rangle + a_2 |{\bf -u} \rangle \otimes
|\downarrow_1\rangle \\ & \to & a_1 | {\bf u}\rangle \otimes
|\uparrow_1\rangle + a_2 |{\bf -u} \rangle \otimes |\uparrow_1\rangle
\, ,
\end{array} 
\end{equation}
generated by the unitary operator
\begin{equation}
  U( t ) = P({\bf u}) \otimes I + P({\bf -u}) \otimes
  \mbox{e}^{-i\theta(t) F} \, ,
\end{equation}
where
\begin{equation}
\theta (t) = \left\{ \begin{array}{ll} t & \mbox{for $0 \leq t \leq
\pi/2$,} \\ \pi/2 & \mbox{for $\pi/2 \leq t \leq \pi$,} \\ 3 \pi/2 - t
& \mbox{for $\pi \leq t \leq 3\pi/2$.}
\end{array} \right.
\end{equation}

We have taken it for granted thus far that a dynamical algorithm makes
selections at time $t$ based only on the evolution of the system up to
that time.  Thus any dynamical algorithm which behaves sensibly,
according to the criteria which we have used so far, will select a
consistent set which includes the Schmidt projections at some time
between $\pi/2$ and $\pi$, since during that interval the projections
appear to describe the result of a completed measurement.  These
projections cannot be consistently extended by projections onto the
initial state $ a_1 | {\bf u}\rangle + a_2 |{\bf -u} \rangle $ and the
orthogonal state $ \overline{a_2} | {\bf u}\rangle - \overline{a_1}
|{\bf -u} \rangle $ at time $3 \pi/2$, so that the algorithm will not
agree with the standard intuition that at time $\pi$ the state of the
system particle has reverted to its initial state.  In particular, if
the particle subsequently undergoes interactions of the form
(\ref{measurementinteraction}) with other environment particles, the
algorithm cannot reproduce the standard description of these later
measurements.  The same problem afflicts the quasi-dynamical algorithm
considered in subsection~\ref{quasi}.

In principle, then, dynamical set selection algorithms of the type
considered so far imply that, following any experiment in which exact
decoherence is followed by exact recoherence and then by a
probabilistic measurement of the recohered state, the standard
quasiclassical picture of the world cannot generally be recovered.  If
the algorithms use an approximate consistency criterion --- as we have
argued is necessary for a realistic algorithm --- then this holds true
for experiments in which the decoherence and recoherence are
approximate.

We know of no experiments of precisely this type.  Several neutron
interferometry experiments have been performed in which one or both
beams interact with an electromagnetic field before
recombination\cite{mds,wb,abr,wbrs,nr,brs,brt,ackokw} and measurement.
In these experiments, though, the electromagnetic field states are
typically superpositions of many different number states, and are
largely unaffected by the interaction, so that (\ref{recoh}) is a poor
model for the process.\footnote{See, for example,
ref. \cite{summhammer} for a review and analysis.}  Still, it seems
hard to take seriously the idea that if a recoherence experiment were
constructed with sufficient care it would jeopardise the
quasiclassicality we observe, and we take the recoherence problem as a
conclusive argument against the general applicability of the
algorithms considered to date.

\section{Retrodictive algorithms}\label{sec:retrodiction}

We have seen that dynamical set selection algorithms which run
forwards in time generally fail to reproduce standard physics.  Can an
algorithm be developed for reconstructing the history of a series of
experiments or, in principle, of the universe?

\subsection{Retrodictive algorithms in the spin model} 

We look first at the spin model with separated interactions and
initial state
\begin{equation} 
|\psi (0) \rangle = |{\bf v}\rangle \otimes |\uparrow_1 \ldots
\uparrow_n \rangle \, ,
\end{equation}
and take the first interaction to run from $t=0$ to $t=1$, the 
second from $t=1$ to $t=2$, and so on.  The final state, in the 
Schr\"odinger picture, is
\begin{equation}
  |\psi(n)\rangle = \sum_{\bf \alpha} \sqrt{p_{\bf \alpha}} |\alpha_n
  {\bf u}_n\rangle \otimes | \beta_1 \ldots \beta_n \rangle \,.
\end{equation}
Here ${\bf \alpha} = \{ \alpha_1 , \ldots , \alpha_{n} \}$ runs over
all strings of $n$ plusses and minuses, we write $ \beta_i = \,
\uparrow )$ if $\alpha_i = 1$ and $ \beta_i = \, \downarrow )$ if
$\alpha_i = -1$, and
\begin{equation} 
  p_{\bf \alpha} = 2^{-n}(1 + \alpha_n \alpha_{n-1} {\bf u}_n. {\bf
  u}_{n-1}) \ldots (1 + \alpha_1 {\bf u_1. u_0}) \,.
\end{equation} 

Consider now a set selection algorithm which begins the selection
process at $t=n$ and works backwards in time, selecting an exactly
consistent set defined by system space Schmidt projections.  The
algorithm thus begins by selecting projections onto the Schmidt states
$|\pm{\bf u}_n\rangle$ at $t=n$.  The classification theorem implies
that any Schmidt projection during the time interval $[n-1,n)$ defines
a consistent and non-trivial extension to the set defined by these
projections. If the algorithm involves a parametrised non-triviality
condition with sufficiently small non-triviality parameter $\delta$,
the next projection will thus be made as soon as the non-triviality
condition is satisfied, which will be at some time $t=n-\Delta t$,
where $\Delta t$ is small.

If a non-triviality condition is not used but the limit DHC is used
instead, then a second projection will be made at $t=n$, but the
normalised path projected states will be the same (to lowest order in
$\Delta t$) as for projection at $t=n-\Delta t$.  The classification
theorem then implies that the only possible times at which further
extensions can consistently be made are $t= n-1, \ldots, 1$ and, if
$\delta$ is sufficiently small and the measurement axes are
non-degenerate, the Schmidt projections at all of these times will be
selected.

In fact, this algorithm gives very similar results whether a
non-triviality condition or the limit DHC is used.  We use the limit
DHC here for simplicity of notation.  Since the Schmidt states at the
end of the $k^{\mbox{\scriptsize th}}$ interaction are $|\pm{\bf
u}_k\rangle$, the histories of the selected set are indexed by strings
$\{ \alpha_1 , \ldots , \alpha_{n+1} \}$ consisting of $n+1$ plusses
and minuses.  The corresponding class operators are defined in terms
of the Heisenberg picture Schmidt projections as
\begin{equation}
  P_H^{\alpha_{n+1}}(n) P_H^{\alpha_{n}}(n) P_H^{\alpha_{n-1}} (n-1)
\ldots P_H^{\alpha_{1}}(1) \, .
\end{equation}
Define $C_{\bf \alpha} = P_H^{\alpha_{n}}(n) \ldots
P_H^{\alpha_{1}}(1)$.  Then
\begin{equation}
\begin{array}{rcll} 
P_H^{\alpha_{n+1}}(n) C_{\bf \alpha} &=& C_{\bf \alpha} & \mbox{if
$\alpha_{n+1} = \alpha_n$,}\\ P_H^{\alpha_{n+1}} (n) C_{\bf \alpha}
&=& 0 & \mbox{if $\alpha_{n+1} = -\alpha_n$,}
\end{array}
\end{equation}  
and to calculate the limit DHC eq.~(\ref{limitDHC}) we note that
eq.~(\ref{dPidenta}) implies that
\begin{equation}
\begin{array}{rcl} 
\lim_{\epsilon \to 0} \epsilon^{-1}P_H^{-\alpha_{n}}(n)
  P_H^{\alpha_{n}}(n-\epsilon) \ldots P_H^{\alpha_{1}}(1) &=&
  P_H^{-\alpha_{n}}(n) \dot P_H^{\alpha_{n}}(n) \ldots
  P_H^{\alpha_{1}}(1) \\ &=& \dot P_H^{\alpha_n}(n)
  P_H^{\alpha_{n}}(n) \ldots P_H^{\alpha_{1}}(1)\\ &=& \dot
  P_H^{\alpha_n}(n) C_{\bf \alpha} \,.
\end{array}
\end{equation}
The complete set of class operators (up to multiplicative constants)
is $\{C_{\bf \alpha}, \dot P_H^+(n) C_{\bf \alpha}\}$ and the set of
normalised histories is therefore
\begin{equation}
  \{ |\alpha_n{\bf u}_n\rangle \otimes |{\bf \alpha}\rangle,
    |-\alpha_n{\bf u}_n\rangle \otimes |{\bf \alpha}\rangle\} \, .
\end{equation}

Of these histories, the first $2^n$ have probabilities $p_{\bf \alpha}
= 2^{-n}(1 + \alpha_n \alpha_{n-1} {\bf u}_n. {\bf u}_{n-1}) \ldots (1
+ \alpha_1 {\bf u_1. u_0})$ and have a simple physical interpretation,
namely that the particle was in direction $\alpha_i {\bf u} _i$ at
time $t=i$, for each $i$ from $1$ to $n$, while the second $2^n$ have
zero probability.  Thus the repeated projections that the algorithm
selects at $t=n$, while non-standard, merely introduce probability
zero histories, which need no physical interpretation.  The remaining
projections reproduce the standard description so that, in this
example, at least, retrodictive algorithms work.  While this is
somewhat encouraging, the algorithm's success here relies crucially on
the simple form of the classification of consistent sets in the spin
model, which in turn relies on a number of special features of the
model.  In order to understand the behaviour of retrodictive
algorithms in more generality, we look next at two slightly more
complicated versions of the spin model.

\subsection{Spin model with perturbed initial state} 

Consider now the spin model with a perturbed initial state
$|\psi\rangle + \gamma|\phi\rangle$.  For generic choices of $\phi$
and $\gamma$, there is no non-trivial exactly consistent set of
Schmidt projections, but it is easy to check that the set selected in
the previous section remains approximately consistent to order
$\gamma$, in the sense that the DHC and limit DHC parameters are $O(
\gamma )$.

This example nonetheless highlights a difficulty with the type of
retrodictive algorithm considered so far.  Some form of approximate
consistency criterion is clearly required to obtain physically
sensible sets in this example.  However, there is no obvious reason to
expect that there should be any parameter $\epsilon$ with the property
that a retrodictive algorithm which requires approximate consistency
(via the limit DHC and DHC) to order $\epsilon$ will select a
consistent set whose projections are all similar to those of the set
previously selected.  The problem is that, given any choice of
$\epsilon$ which selects the right projections at time $n$, the next
projections selected will be at time $(n-1) + O ( \gamma )$ rather
than at precisely $t = n-1$.  The level of approximate consistency
then required to select projections at times near $n-2$, $n-3$, and so
forth, depends on the projections already selected, and so depends on
$\gamma$ only indirectly and in a rather complicated way.

We expect that, for small $\gamma$ and generic $\phi$, continuous
functions $\epsilon_k(\gamma, \phi )$ exist with the properties that
$\epsilon_k (\gamma , \phi) \rightarrow 0 $ as $\gamma \rightarrow 0$
and that some approximation to the set previously selected will be
selected by a retrodictive algorithm which requires approximate
consistency to order $\epsilon_k(\gamma, \phi )$ for the
$k^{\mbox{\scriptsize th}}$ projection.  Clearly, though, since the
aim of the set selection program is to replace model-dependent
intuition by a precise algorithmic description, it is rather
unsatisfactory to have to fine-tune the algorithm to fit the model in
this way.

\subsection{Delayed choice spin model}  

We now return to considering the spin model with an unperturbed
initial state and look at another shortcoming.  The interaction of the
system particle with each successive environment particle takes the
form of a spin measurement interaction in which the axis of each
measurement, $\{ {\bf u}_i \}$, is fixed in advance.  This is a
sensible assumption when modelling a natural system-environment
coupling, such as a particle propagating past a series of other
particles.  As a model of a series of laboratory experiments, however,
it is unnecessarily restrictive.  We can model experiments with an
element of delayed choice simply by taking the axis $\{ {\bf u}_i \}$
to depend on the outcome of the earlier measurements.

If we do this, while keeping the times of the interactions fixed and
non-overlapping, the measurement outcomes can still be naturally
described in terms of a consistent set built from Schmidt projections
onto the system space at times $t=1,2,\ldots n$, so long as both the
Schmidt projections and the consistent set are defined to be
appropriately branch-dependent.  Thus, let
\begin{equation} 
|\psi (0) \rangle = |{\bf v}\rangle \otimes |\uparrow_1 \ldots
\uparrow_n \rangle \,
\end{equation} 
be the initial state and let $P_H^{\alpha_{1}}(1)$, for $\alpha_1 =
\pm$, be the Schmidt projections onto the system space at time $t=1$.
We define a branch-dependent consistent set in which these projections
define the first branches and consider independently the evolution of
the two states $ P_H^{+} (1) | \psi(0) \rangle $ and $ P_H^{-} (1) |
\psi(0) \rangle $ between $t=1$ and $t=2$.  These evolutions take the
form of measurements about axes ${\bf u}_{2; \alpha_1 }$ which depend
on the result of the first measurement.  At $t=2$ the second
measurements are complete, each branch splits again, and the
subsequent evolutions of the four branches now depend on the results
of the first two measurements.  Similar splittings take place at each
time from $1$ to $n$, so that the axis of the $m^{\mbox{\scriptsize
th}}$ measurement in a given branch, ${\bf u}_{m; \alpha_{m-1},
\ldots, \alpha_1 } $, depends on the outcomes $\alpha_{m-1} , \ldots ,
\alpha_1$ of the previous $(m-1)$ measurements.  Thus, the evolution
operator describing the $m^{\mbox{\scriptsize th}}$ interaction is
\begin{eqnarray*}
\lefteqn{ V_m( t ) = } \\ & \displaystyle \sum_{\alpha_{m-1}, \ldots ,
\alpha_1} & \{ P({\bf u}_{m; \alpha_{m-1}, \ldots, \alpha_1 } )
\otimes P_1 ( \beta_1 ) \otimes \ldots \otimes P_{m-1} ( \beta_{m-1} )
\otimes I_m \otimes \ldots \otimes I_n \, + \\ & & \hspace{-.7ex} P( -
{\bf u}_{m; \alpha_{m-1}, \ldots, \alpha_1} ) \otimes P_1 ( \beta_1)
\otimes \ldots \otimes P_{m-1} ( \beta_{m-1}) \otimes
\mbox{e}^{-i\theta_m (t) F_m } \otimes I_{m+1} \otimes \ldots \otimes
I_n \} \, .
\end{eqnarray*} 
Again we take $\beta_i = \, \uparrow$ if $\alpha_i = +$ and $\beta_i =
\, \downarrow$ if $\alpha_i = - $.  The full evolution operator is
\begin{equation}
  U(t) = V_n (t) \ldots V_1 (t) \, .
\end{equation} 
During the interval $(m-1,m)$ we consider the Schmidt decompositions
on each of the $2^{m-1}$ branches defined by the states
\begin{eqnarray*} 
  && U(t) P_H^{\alpha_{m-1} ; \alpha_{m-2} , \ldots, \alpha_1} (m-1)
  \ldots P_H^{\alpha_1} (1) | \psi(0) \rangle \\ &=& V_m(t) [ P(
  \alpha_{m-1} {\bf u}_{m-1; \alpha_{m-2} , \ldots, \alpha_1}) \ldots
  P( \alpha_1 {\bf u}_1 ) | {\bf v} \rangle] \otimes |\beta_1 \ldots
  \beta_{m-1} \uparrow_m \ldots \uparrow_n\rangle\,
\end{eqnarray*} 
with $\alpha_1 , \ldots , \alpha_{m-1}$ independently running over the
values $\pm$.  Here
\begin{equation} 
P_H^{\alpha_m ; \alpha_{m-1}, \ldots , \alpha_1} (t) = U^{\dagger} (t)
P( \alpha_{m} {\bf u}_{m; \alpha_{m-1} , \ldots, \alpha_1}) \otimes I
U(t) \, ,
\end{equation} 
that is, the Heisenberg picture projection operator onto the
branch-dependent axis of measurement.  The branches, in other words,
are defined by the branch-dependent Schmidt projections at times from
$1$ to $m-1$.

It is not hard, thus, to find a branch-dependent consistent set, 
built from the branch-dependent Schmidt projections at times $1$
through to $n$, which describes the delayed-choice spin model
sensibly.\footnote{This sort of branch-dependent Schmidt decomposition
could, of course, be considered in the original spin model, where all
the axes of measurement are predetermined, but would not affect the
earlier analysis, since the Schmidt projections in all branches are
identical.}  However, since the retrodictive algorithms considered so
far rely on the existence of a branch-independent set defined by the
Schmidt decompositions of the original state vector, they will not
generally reproduce this set (or any other interesting set).
Branch-dependent physical descriptions, which are clearly necessary in
quantum cosmology as well as in describing delayed-choice experiments,
appear to rule out the type of retrodictive algorithm we have
considered so far.

\section{Branch-dependent algorithms} \label{sec:branchdep}

The algorithms we have considered so far do not allow for
branch-dependence, and hence cannot possibly select the right set in
many physically interesting examples.  We have also seen that it is
hard to find good Schmidt projection selection algorithms in which the
projections selected at any time depend only on the physics up to that
time, and that the possibility of recoherence rules out the existence
of generally applicable algorithms of this type.

This suggests that \emph{retrodictive} branch-dependent algorithms
should be considered.  Such algorithms, however, seem generally to
require more information than is contained in the evolution of the
quantum state.  In the delayed-choice spin model, for example, it is
hard to see how the Schmidt projections on the various branches,
describing the delayed-choice measurements at late times, could be
selected by an algorithm if only the entire state $\psi(t)$ --- summed
over all the branches --- is specified.

The best, we suspect, that can be hoped for in the case of the
delayed-choice spin model is an algorithm which takes all the final
branches, encoded in the $2^n$ states $| \pm {\bf v} \rangle \otimes |
\beta_1 \ldots \beta_n \rangle$, where each of the $\beta_i$ is one of
the labels $\uparrow$ or $\downarrow$, and attempts to reconstruct the
rest of the branching structure from the dynamics.

One possibility, for example, is to work backwards from $t=n$, and at
each time $t$ search through all subsets $Q$ of branches defined at
that time, checking whether the sum $| \psi^Q (t) \rangle$ of the
corresponding states at time $t$ has a Schmidt decomposition with the
property that the Schmidt projections, applied to $| \psi^Q(t)
\rangle$, produce (up to normalisation) the individual branch states.
If so, the Schmidt projections are taken to belong to the selected
branch-dependent consistent set, the corresponding branches are
unified into a single branch at times $t$ and earlier, and the state
corresponding to that branch at time $t'$ is taken to be $ U(t')
U(t)^{\dagger} | \psi^Q (t) \rangle$, where $U$ is the evolution
operator for the model.  Clearly, though, by specifying the final
branch states we have already provided significant information ---
arguably most of the significant information --- about the physics of
the model.  Finding algorithmic ways of supplying the branching
structure of a natural consistent set, given all of its final history
states, may seem a relatively minor accomplishment.  It would
obviously be rather more useful, though, if the final history states
themselves were specified by a simple rule.  For example, if the
system and environment Hilbert spaces are both of large dimension, the
final Schmidt states would be natural candidates.  It would be
interesting to explore these possibilities in quantum cosmology.

\section{Conclusions}

John Bell, writing in 1975, said of the continuing dispute about
quantum measurement theory that it ``is not between people who
disagree on the results of simple mathematical manipulations.  Nor is
it between people with different ideas about the actual practicality
of measuring arbitrarily complicated observables.  It is between
people who view with different degrees of concern or complacency the
following fact: so long as the wave packet reduction is an essential
component, and so long as we do not know exactly when and how it takes
over from the Schr\"odinger equation, we do not have an exact and
unambiguous formulation of our most fundamental physical
theory.''\cite{bell:hepp}

New formulations of quantum theory have since been developed, and the
Copenhagen interpretation itself no longer dominates the debate quite
as it once did.  The language of wave packet reduction, in particular,
no longer commands anything approaching universal acceptance ---
thanks in large part to Bell's critiques.  But the fundamental dispute
is still, of course, very much alive, and Bell's description of the
dispute still essentially holds true.  Many approaches to quantum
theory rely, at the moment, on well-developed intuition to explain,
case by case, what to calculate in order to obtain a useful
description of the evolution of any given physical system.  The
dispute is not over whether those calculations are correct, or even as
to whether the intuitions used are helpful: generally, both are.  The
key question is whether we should be content with these successes, or
whether we should continue to seek to underpin them by an exact and
unambiguous formulation of quantum theory.

Consensus on this point seems no closer than it was in 1975.  Many
physicists take the view that we should not ever expect to find a
complete and mathematically precise theory of nature, that nature is
simply more complex than any mathematical representation.  If so, some
would argue, present interpretations of quantum theory may well
represent the limit of precision attainable: it may be impossible, in
principle, to improve on imprecise verbal prescriptions and intuition.
On the other hand, this doubt could be raised in connection with any
attempt to tackle any unsolved problem in physics.  Why, for example,
should we seek a unified field theory, or a theory of turbulence, if
we decide a priori not to look for a mathematically precise
interpretation of quantum theory?  Clearly, too, accepting the
impossibility of finding a complete theory of nature need not imply
accepting that any definite boundary to precision will ever be
encountered.  One could imagine, for example, that every technical and
conceptual problem encountered can eventually be resolved, but that
the supply of problems will turn out to be infinite.  And many
physicists, of course, hope or believe that a complete and compelling
theory of nature will ultimately be found, and so would simply reject
the initial premise.

Complete agreement on the desiderata for formulations of quantum
theory thus seems unlikely.  But it ought to be possible to agree
whether any given approach to quantum theory actually does supply an
exact formulation and, if not, what the obstacles might be.  Our aim
in this paper has been to help bring about such agreement, by
characterising what might constitute a precise formulation of some of
the ideas in the decoherence and consistent histories literature, and
by explaining how hard it turns out to be to supply such a
formulation.

Specifically, we have investigated various algorithms that select one
particular consistent set of histories from among those defined by the
Schmidt decompositions of the state, relative to a fixed
system-environment split.  We give examples of partial successes.
There are several relatively simple algorithms which give physically
sensible answers in particular models, and which we believe might
usefully be applied elsewhere.  We have not, though, found any
algorithm which is guaranteed to select a sensible consistent set when
both recoherence and branch-dependent system-environment interactions
are present.

Our choice of physical models is certainly open to criticism.  The
spin model, for example, is a crudely simplistic model of real world
decoherence processes, which supposes both that perfect correlations
are established between system and environment particles in finite
time and that these interactions do not overlap.  We would not claim,
either, that the delayed-choice spin model necessarily captures any of
the essential features of the branching structure of quasiclassical
domains, though we would be very interested to know whether it might.
We suspect that these simplifications should make it easier rather
than harder to find set selection algorithms in the models, but we
cannot exclude the possibility that more complicated and realistic
models might prove more amenable to set selection.

The type of mathematical formulation we have sought is, similarly,
open to criticism.  We have investigated what seem particularly
interesting classes of Schmidt projection set selection algorithms,
but there are certainly others which may be worth exploring.  There
are also, of course, other mathematical structures relevant to
decoherence apart from the Schmidt decomposition, and other ways of
representing historical series of quantum events than through
consistent sets of histories.

Our conclusion, though, is that it is extraordinarily hard to find a
precise formulation of non-relativistic quantum theory, based on the
notions of quasiclassicality or decoherence, that is able to provide a
probabilistic description of series of events at different points in
time sufficiently rich to allow our experience of real world physics
to be reconstructed.  The problems of recoherence and of
branch-dependent system-environment interaction, in particular, seem
sufficiently serious that we doubt that the ideas presented in the
literature to date are adequate to provide such a formulation.
However, we cannot claim to have exhaustively investigated every
possibility, and we would like to encourage sceptical readers to
improve on our attempts.

\section*{Acknowledgments}

We are grateful to Fay Dowker and Trevor Samols for several helpful
discussions and to Philip Pearle for raising the question of the
implications of recoherence for set selection algorithms at an early
stage of this work and for other valuable comments.  A.K. was
supported by a Royal Society University Research Fellowship; J.M. by
the United Kingdom Engineering and Physical Sciences Research Council.

\end{document}